\pdfoutput=1
\documentclass[useAMS,usenatbib]{mn2e}
\usepackage{graphicx} 
\usepackage{amsmath} 

\title[The  STEP Survey]{STEP: The VST survey of the SMC and the
  Magellanic Bridge. I. Overview and first results.\thanks{This work is based on 
INAF-VST guaranteed observing time under ESO programmes 089.D-0258; 090.D-0172}}

\author[V. Ripepi et al.]
{V. Ripepi,$^{1}$\thanks{E-mail: ripepi@oacn.inaf.it}
 M. Cignoni,$^{2,3}$ M. Tosi,$^{3}$ M. Marconi,$^{1}$ I. Musella,$^{1}$  A. Grado,$^{1}$  
\newauthor{L. Limatola,$^{1}$ G. Clementini,$^{3}$ E. Brocato,$^{4}$ M. Cantiello,$^{5}$ M. Capaccioli,$^{6}$} 
 \newauthor{E. Cappellaro,$^{7}$ M-R. L. Cioni,$^{8,9}$ F. Cusano,$^{3}$ M. Dall'Ora,$^{1}$  J. S. Gallagher III,$^{10}$}
\newauthor{E. K. Grebel,$^{11}$ A. Nota,$^{2,12}$ F. Palla,$^{13}$D. Romano,$^{3}$ G. Raimondo,$^{5}$}
\newauthor{E. Sabbi,$^{2}$ F. Getman,$^{1}$ N. R. Napolitano,$^{1}$
  P. Schipani,$^{1}$ S. Zaggia$^{7}$}
\\
\\
$^{1}$ INAF-Osservatorio Astronomico di Capodimonte, Via Moiariello
 16, 80131, Naples, Italy \\
$^{2}$ Space Telescope Science Institute, 3700 San Martin
Drive, Baltimore, USA\\
$^{3}$ INAF-Osservatorio Astronomico di Bologna, via Ranzani 1,
Bologna, Italy \\ 
$^{4}$ INAF-Osservatorio Astronomico di Roma, Via Frascati 33, 00040, Monte Porzio Catone (RM), Italy \\
$^{5}$ INAF-Osservatorio Astronomico di Teramo, Via M. Maggini, 64100 Teramo, Italy \\
$^{6}$ Dipartimento di Fisica, Universit\`a ``Federico II'', Naples, Italy \\
$^{7}$ INAF-Osservatorio Astronomico di Padova, Vicolo dell'Osservatorio 5,  35122 Padova, Italy \\
$^{8}$ University of Hertfordshire, Physics Astronomy and
 Mathematics, Hatfield AL10 9AB, UK \\
$^{9}$ Leibnitz-Institut f\"{u}r Astrophysik Potsdam, An der
Sternwarte 16, 14482 Potsdam, Germany \\
$^{10}$ Department of Astronomy, University of Wisconsin-Madison, 5534 Sterling, 475 North Charter Street, Madison, WI 53706, USA\\
$^{11}$ Astronomisches Rechen-Institut, Zentrum f\"{u}r Astronomie der
Universit\"{a}t Heidelberg, \\
M\"{o}nchhofstr 12-14, D-69120 Heidelberg, Germany\\
$^{12}$ European Space Agency, Research and Scientific Support Department, Baltimore, MD, USA\\
$^{13}$ INAF-Osservatorio Astroﬁsico di Arcetri, Largo E. Fermi 5, 50125 Firenze, Italy\\
}

\begin{document}

\date{}

\pagerange{\pageref{firstpage}--\pageref{lastpage}} \pubyear{2002}

\maketitle

\label{firstpage}

\begin{abstract}
STEP (the {\bf S}MC in {\bf T}ime: {\bf E}volution of a {\bf
  P}rototype interacting late-type dwarf galaxy) is a Guaranteed Time
Observation survey being performed at the VST (the ESO VLT Survey
Telescope).  STEP will image an area of 74 deg$^2$ covering the main
body of the Small Magellanic Cloud (32 deg$^2$), the Bridge that
connects it to the Large Magellanic Cloud (30 deg$^2$) and a small
part of the Magellanic Stream (2 deg$^2$). Our $g,r,i,H_{\alpha}$
photometry is able to resolve individual stars down to magnitudes well
below the main-sequence turnoff of the oldest populations.  In this
first paper we describe the observing strategy, the photometric
techniques, and the upcoming data products of the survey. We also
present preliminary results for the first two fields for which data
acquisition is completed, including some detailed analysis of the two
stellar clusters IC\,1624 and NGC\,419.

\end{abstract}


\section{Introduction}

Dwarf galaxies in the Local Group provide a distinct  laboratory for
studying and testing galaxy formation theories and cosmology. 
To understand the history of formation and evolution of galaxies, as
well as their interactions with the environment, we need to study the
3D structure and the distributions of age, chemical abundance and
kinematics of both stellar and gaseous components. We also need to
obtain a global picture of the star formation history (SFH),
accounting for both the intrinsic evolution and the effects of
interactions with neighbouring systems.  This is best achieved in
nearby galaxies, where individual stars can be resolved
down to faint absolute magnitudes, thus probing stellar masses from
high to low and ages from young to old. In closer systems, the
stellar populations can be characterised in exquisite detail, with
deep photometry and spectroscopy, and their SFHs can be reliably
derived over the whole Hubble time, from the most recent epochs back
to the earliest ones \citep[see e.g.,] [and references therein]{tolstoy09}.

The Small Magellanic Cloud \citep[SMC, $M_V=-16.8\pm0.2$ mag; $D_{\sun}
\sim 64$ Kpc;][]{devau1991, Udalski1999} is the closest dwarf galaxy of late
morphological type, hence the ideal target for detailed studies of
the properties of this most common class of galaxies. Its mass
\citep[between 1 and 5 $\times 10^9$ M$_{\sun}$, e.g.,][and references
therein]{kallivayalil06,Bekki2009} is at the upper limit of the typical 
late-type dwarf masses. Its high gas content
\citep[M$_{HI} =4 \times 10^8$ M$_{\sun}$][]{Bruns2005} and
low present-day metallicity \citep[Z=0.004, see e.g.,][]{Russel1989} are also common features 
of dwarf irregulars (dIrr) and starburst dwarfs. 

The old populations in dIrrs may be representative of the stellar content
of the early, gas--rich galaxies that are believed to have 
contributed significantly to the build-up of more massive 
galaxies \citep[e.g.,][]{Delucia2008,Cooper2010}
more than 9 Gyr ago.  These old stellar populations are
quite a bit more metal-poor than the comparatively high
present-day SMC metallicity of Z=0.004 \citep[see e.g.,][]{Kunth2000}.


As a member of the nearest group of interacting galaxies, the SMC is
also an ideal benchmark to study the effects of tidal interactions on
galaxy evolution. In fact there are clear signatures that the SMC is
interacting with its neighbours, the Large Magellanic Cloud (LMC) and
the Milky Way (MW).  In particular, the Magellanic Clouds (MCs) are
connected by an HI-dominated gaseous Bridge
\citep{Hindman1963,Irwin1985,harris07} that, like the Magellanic
Stream, may result from their mutual gravitational effects and/or the
influence of the MW.  The SMC Wing, the portion of the SMC main body
protruding asymmetrically towards the LMC \citep{Shapley1940}, may
also be the result of tidal forces. Also, the bar of the SMC, which
stands out when using young populations as tracers, has a highly
asymmetric and elongated structure with its north--eastern part closer
to us than its south--western part \citep{Haschke2012}. To
  complicate matters further, remarkable line-of-sight depth
  variations are well established in the SMC. For instance, 
\citet[][]{hatzi89}, using horizontal branch (HB) and red clump
  (RC) stars, found that the northeastern periphery of the galaxy (at
  more than 2 kpc from the optical centre) suffers an average depth of
  17 kpc with a maximum of 23 kpc, while the southwest region shows a depth of
  about 10 kpc on average. More recently, \citet[][]{subra09} found a
  depth between 0.67 kpc and 9.53 kpc, with an increase near
  the optical centre. Finally, \citet[][]{glatt08b} found depths
  between 10 and 17 kpc using the RC of several populous star
  clusters.

 State-of-the-art computations based on Hubble Space Telescope
(HST) data \citep{kallivayalil13} assume that the Clouds are bound
to each other, with the SMC on an elliptical orbit around the LMC. It
was estimated that the frequency of MC-like systems in the Universe is
only a few percent
\citep[e.g.,][]{robotham12} and they may have entered the MW potential
only recently.

The Bridge holds important clues about the most recent (the last
$\sim$100 Myr) interaction between the Clouds,
its own formation history and possible origin. Intermediate-age and
old stars, probably stripped by either Cloud, may also be present in
the Bridge \citep{bagheri13, noel13} and would provide additional
information about its origin.

\begin{figure}
\includegraphics[width=8.5cm]{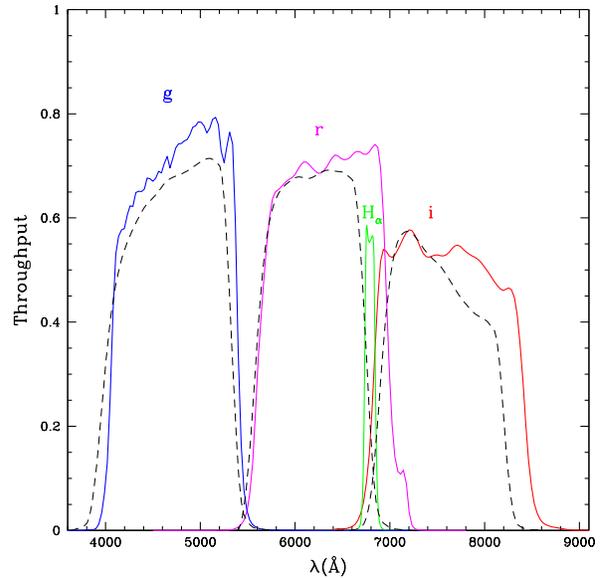}
\caption{Transmittance of the filters used in STEP (solid lines). The contribution of 
the CCD is included. The dashed curves represent the transmission 
curves for the corresponding SDSS filters (properly rescaled to 
favour the comparison).}
\label{figureFilter}
\end{figure}

\begin{figure*}
\includegraphics[width=18cm]{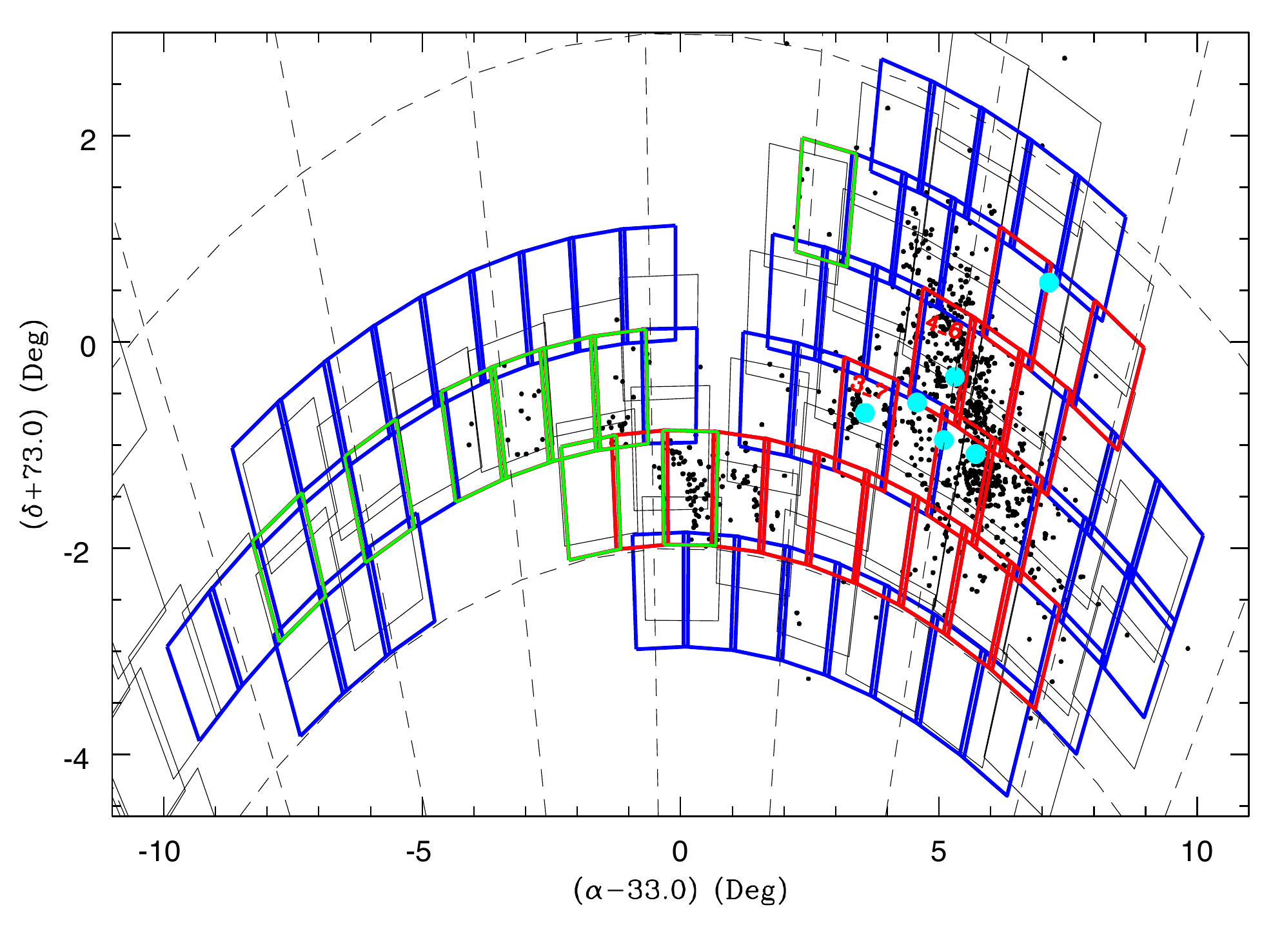}
\caption{Map of STEP's tiles (the two  tiles centred in the 
  direction of the Magellanic Stream are out of the figure). To 
  highlight the location of the SMC body and of part of the 
  Bridge, black dots indicate the position of known star clusters and 
  associations \citep[according to][]{bic08}. The thick boxes correspond to 
  the 1 deg$^2$ FoV of the  VST tiles. Red boxes 
  represent tiles whose observations are completed, green boxes those 
  with completed time-series photometry, and blue boxes the remaining 
  ones.  To find a correspondance with Tab.~\ref{tabFields}, note that 
  the tiles from south to north are those with prefix ``1\_'' and 
  ``6\_'', respectively. From west to east the first and last tiles 
  have suffix ``\_1'' and ``\_21'', respectively.  For comparison,
  grey thin boxes show the VMC tiles, whereas 
  the HST fields described by \citet{sabbi09} are the small cyan 
  filled circles (note that the true size of HST fields is significantly 
smaller). The two tiles analysed in this work (tiles 3\_7 and 4\_6) are labelled.}
\label{mapStep}
\end{figure*}

A wealth of data on the SMC are available in the literature, although
not as complete as for its bigger companion, the LMC. Yet, we are
still missing crucial information for our understanding of how the SMC has formed and
evolved. For these reasons, a long-term international project has been
set up with the aim of studying in detail the stellar populations, the
structure and the evolution of the SMC in space and time, thanks to
the exploitation of high performance telescopes, such as the HST, the VLT 
(Very Large Telescope)  and other 10 m class telescopes.  The project aims at collecting both
photometric and spectroscopic data to study the (variable and
non-variable) stellar populations of the SMC, from the oldest to the
youngest objects and derive their chemical abundances, spatial
distribution and kinematics. In particular, HST is providing high
spatial resolution photometry of selected SMC fields, along with old and young
clusters \citep[P.I.s J. Gallagher and A. Nota, see e.g.,][]{nota06,
  sabbi07, glatt08a, glatt08b, glatt09, glatt11,sabbi09,
  cignoni10b,cignoni11,cignoni12,cignoni13}, whereas VLT spectroscopy
with FORS in the CaII triplet (P.I. E. Grebel) allows to study the
age-metallicity relation of the SMC clusters \citep{Kayser2008}.

\begin{table}
\caption{Characteristics of the filters used in the STEP
  survey.
}
\label{tabFilters}
\begin{center}
\begin{tabular}{cccccc}
\hline
\hline 
\noalign{\smallskip} 
Filter & $\lambda_{\rm mean}$ & $\lambda_{\rm eff}^a$ & $\lambda_{\rm min}$ &$\lambda_{\rm max}$  & $\Delta \lambda_{\rm eff}^b$   \\
\noalign{\smallskip}
 & \AA & \AA&\AA &\AA & \AA  \\
\noalign{\smallskip}
\hline
\noalign{\smallskip} 
$g$             &  4760 & 4680 & 3913  &  5573  &  1203      \\
$r$              & 6326  &6242  &5405  &  7245   & 1314       \\
$i$               & 7599 & 7509 & 6580 &   8741   & 1464   \\
$H\alpha$   & 6789 & 6789 & 6675  &  6910  &  119     \\
\noalign{\smallskip}
\hline
\noalign{\smallskip}
\end{tabular}
\end{center}
$^a$ central filter wavelength after the  transmission curve 
  was convolved with the spectrum of Vega\\
$^b$ effective filter width, equivalent to the horizontal size of a
rectangle with height equal to the maximum transmission curve and with the
same area of the one covered by the filter transmission curve\\
\end{table}


In this context, the scope of this paper is to present the first
results from the VST (VLT Survey Telescope) survey STEP (The {\bf S}MC
in {\bf T}ime: {\bf E}volution of a {\bf P}rototype interacting
late-type dwarf galaxy, PI V. Ripepi). 
We describe the STEP survey in detail in Section 2, while the data
reduction is reported in Section 3. The resulting CMDs for two fields
are discussed in Section 4, an example of analysis of star clusters
is given in Section 5. A brief summary closes the paper.

\begin{table*}
\caption{STEP field centres and status of observations as of 2014, January. A ``TS'' in the columns
  ``MODE'' means that the field was observed in time-series mode. In the ``NOTES'' columns: C=Completed; S=Started. 
The two planned tiles in the direction of the Magellanic Stream
are not included here as their centres coincides with those reported in Tab. A.4 of \citet{Cioni11}.
}
\label{tabFields}
\begin{center}
\begin{tabular}{ccccccccccc}
\hline
\hline 
\noalign{\smallskip} 
Field & RA (J2000) & DEC (J2000) & MODE & NOTES & &Field & RA (J2000) & DEC (J2000) & MODE & NOTES \\
\noalign{\smallskip}
\hline
\noalign{\smallskip} 
   1\_3  & 00:32:11.376  &  -75:24:00.360   &      &      & ~~~~~~~~   &  3\_16  & 03:21:58.800  &  -73:24:58.680   &      &  S  \\
   1\_4  & 00:47:17.760  &  -75:24:00.360   &      &      & ~~~~~~~~   & 3\_17  & 03:35:22.584  &  -73:24:58.680   &  TS  &  C  \\
   1\_5  & 01:02:24.120  &  -75:24:00.360   &      &      & ~~~~~~~~   & 3\_18  & 03:48:46.368  &  -73:24:58.680   &      &  S  \\
   1\_6  & 01:17:30.504  &  -75:24:00.360   &      &      & ~~~~~~~~   &  3\_19  & 04:02:10.152  &  -73:24:58.680   &  TS &  C  \\
   1\_7  & 01:32:36.888  &  -75:24:00.360   &      &      & ~~~~~~~~   & 3\_20  & 04:15:33.960  &  -73:24:58.680   &      &     \\
   1\_8  & 01:47:43.272  &  -75:24:00.360   &      &      & ~~~~~~~~   & 3\_21  & 04:28:57.744  &  -73:24:58.680   &      &    \\
   1\_9  & 02:02:49.632  &  -75:24:00.360   &      &      & ~~~~~~~~   & 4\_1  & 00:00:38.544  &  -72:25:27.480   &      &     \\
  1\_10  & 02:17:56.016  &  -75:24:00.360   &      &     &  ~~~~~~~~  & 4\_2  & 00:13:19.584  &  -72:25:27.480   &      &    \\
   2\_2  & 00:15:40.368  &  -74:24:29.520   &      &      & ~~~~~~~~   & 4\_3  & 00:26:00.648  &  -72:25:27.480   &      &  S  \\
   2\_3  & 00:29:52.224  &  -74:24:29.520   &      &  C  &  ~~~~~~~~  &  4\_4  & 00:38:41.712  &  -72:25:27.480   &      &  C  \\
   2\_4  & 00:44:04.104  &  -74:24:29.520   &      &  C  &  ~~~~~~~~  &  4\_5  & 00:51:22.752  &  -72:25:27.480   &      &  C  \\
   2\_5  & 00:58:15.984  &  -74:24:29.520   &      &  C  &  ~~~~~~~~  &  4\_6  & 01:04:03.816  &  -72:25:27.480   &      &  C  \\
   2\_6  & 01:12:27.864  &  -74:24:29.520   &      &  S  &  ~~~~~~~~   &  4\_7  & 01:16:44.856  &  -72:25:27.480   &      &  S  \\
   2\_7  & 01:26:39.744  &  -74:24:29.520   &      &  C  &  ~~~~~~~~  &  4\_8   & 01:29:25.920  &  -72:25:27.480   &      &    \\
  2\_8   & 01:40:51.624  &  -74:24:29.520   &      &  C  &  ~~~~~~~~  & 4\_9   & 01:42:06.960  &  -72:25:27.480   &      &  S  \\
  2\_9   & 01:55:03.504  &  -74:24:29.520   &      &  C  &  ~~~~~~~~  & 4\_12  & 02:20:10.128  &  -72:25:27.480   &      &    \\
  2\_10  & 02:09:15.384  &  -74:24:29.520   &  TS  &  C  & ~~~~~~~~   &   4\_13  & 02:32:51.168  &  -72:25:27.480   &      &    \\
  2\_11  & 02:23:27.264  &  -74:24:29.520   &      &  C   &~~~~~~~~   & 4\_14  & 02:45:32.232  &  -72:25:27.480   &      &    \\
  2\_12  & 02:37:39.120  &  -74:24:29.520   &      &  C   &~~~~~~~~   & 4\_15  & 02:58:13.272  &  -72:25:27.480   &      &    \\
  2\_16  & 03:34:26.640  &  -74:24:29.520   &      &      &~~~~~~~~    &4\_16  & 03:10:54.336  &  -72:25:27.480   &      &    \\
  2\_17  & 03:48:38.520  &  -74:24:29.520   &      &      &~~~~~~~~     &  4\_17  & 03:23:35.400  &  -72:25:27.480   &      &    \\
  2\_18  & 04:02:50.400  &  -74:24:29.520   &      &      &~~~~~~~~     & 4\_18  & 03:36:16.440  &  -72:25:27.480   &      &    \\
   3\_1  & 00:01:01.992  &  -73:24:58.680   &      &      &~~~~~~~~      & 4\_19  & 03:48:57.504  &  -72:25:27.480   &      &    \\
   3\_2  & 00:14:25.776  &  -73:24:58.680   &      &      &~~~~~~~~      &  4\_20  & 04:01:38.544  &  -72:25:27.480   &      &    \\
   3\_3  & 00:27:49.560  &  -73:24:58.680   &      &      &~~~~~~~~      &  5\_3  & 00:24:23.304  &  -71:25:56.640   &      &  C  \\
   3\_4  & 00:41:13.368  &  -73:24:58.680   &      &  C   &~~~~~~~~     &  5\_4  & 00:36:26.160  &  -71:25:56.640   &      &  S  \\
   3\_5  & 00:54:37.152  &  -73:24:58.680   &      &  C   &~~~~~~~~     &   5\_5  & 00:48:28.992  &  -71:25:56.640   &      &  C  \\
   3\_6  & 01:08:00.936  &  -73:24:58.680   &      &  S   &~~~~~~~~      &  5\_6  & 01:00:31.824  &  -71:25:56.640   &      &    \\
   3\_7  & 01:21:24.720  &  -73:24:58.680   &      &  C   &~~~~~~~~      &   5\_7  & 01:12:34.680  &  -71:25:56.640   &      &    \\
  3\_8   & 01:34:48.504  &  -73:24:58.680   &      &       &~~~~~~~~        &  5\_8  & 01:24:37.512  &  -71:25:56.640   &      &    \\
  3\_9   & 01:48:12.288  &  -73:24:58.680   &      &  S   &~~~~~~~~       &   5\_9  & 01:36:40.344  &  -71:25:56.640   &  TS  &  C  \\
  3\_11  & 02:14:59.856  &  -73:24:58.680   &      &      &~~~~~~~~        &  6\_4  & 00:34:24.312  &  -70:26:25.440   &      &    \\
  3\_12  & 02:28:23.664  &  -73:24:58.680   &  TS  &  C   &~~~~~~~~     &    6\_5  & 00:45:52.800  &  -70:26:25.440   &      &  S  \\
  3\_13  & 02:41:47.448  &  -73:24:58.680   &  TS  &  C   &~~~~~~~~     &   6\_6   & 00:57:21.264  &  -70:26:25.440   &      &     \\
  3\_14  & 02:55:11.232  &  -73:24:58.680   &  TS  &  C   &~~~~~~~~    &  6\_7  & 01:08:49.752  &  -70:26:25.440   &      &    \\
  3\_15  & 03:08:35.016  &  -73:24:58.680   &  TS  &  C   &~~~~~~~~    &  6\_8  & 01:20:18.216  &  -70:26:25.440   &      &    \\
\noalign{\smallskip}
\hline
\noalign{\smallskip}
\end{tabular}
\end{center}
\end{table*}

\section{The STEP Survey}
\label{stepSurvey}

The STEP survey is based on VST Guaranteed Time
Observations (GTO; PI V.Ripepi). 
The survey aims at performing the first deep and
homogeneous optical photometric monitoring of the entire SMC body as well
as of the Bridge. The VST \citep{VST} is an ideal instrument for a 
homogeneous survey of the properties of  the stellar populations present in
the SMC components (main body, Wing, Bridge, outskirts), thanks to its wide
Field--of--View (FoV, 1 deg$^2$) and a magnitude limit optimal for the SMC
distance. It provides the most natural complement to HST imaging,
which reaches much fainter magnitudes with high spatial 
resolution, but over very small fields of view.



Some of us (V. Ripepi, M.R. Cioni., G. Clementini and M. Marconi) are deeply
involved in the VMC@VISTA European Southern Observatory (ESO)  Public survey (P.I: M.R. Cioni) that
is  achieving $YJK_\mathrm{s}$ photometry of the Magellanic System (LMC, SMC,
Bridge, Stream, 184 deg$^2$ at Ks=20.3 mag). STEP and VMC@VISTA are
designed to complement each other: the target fields of the two surveys
overlap completely in the SMC and Bridge region, and will provide,
when completed, homogeneous
$g,r,i,H_{\alpha},Y,J,K_\mathrm{s}$ photometry for
millions of stars in the SMC and the Bridge.

\begin{figure}
\includegraphics[width=9cm]{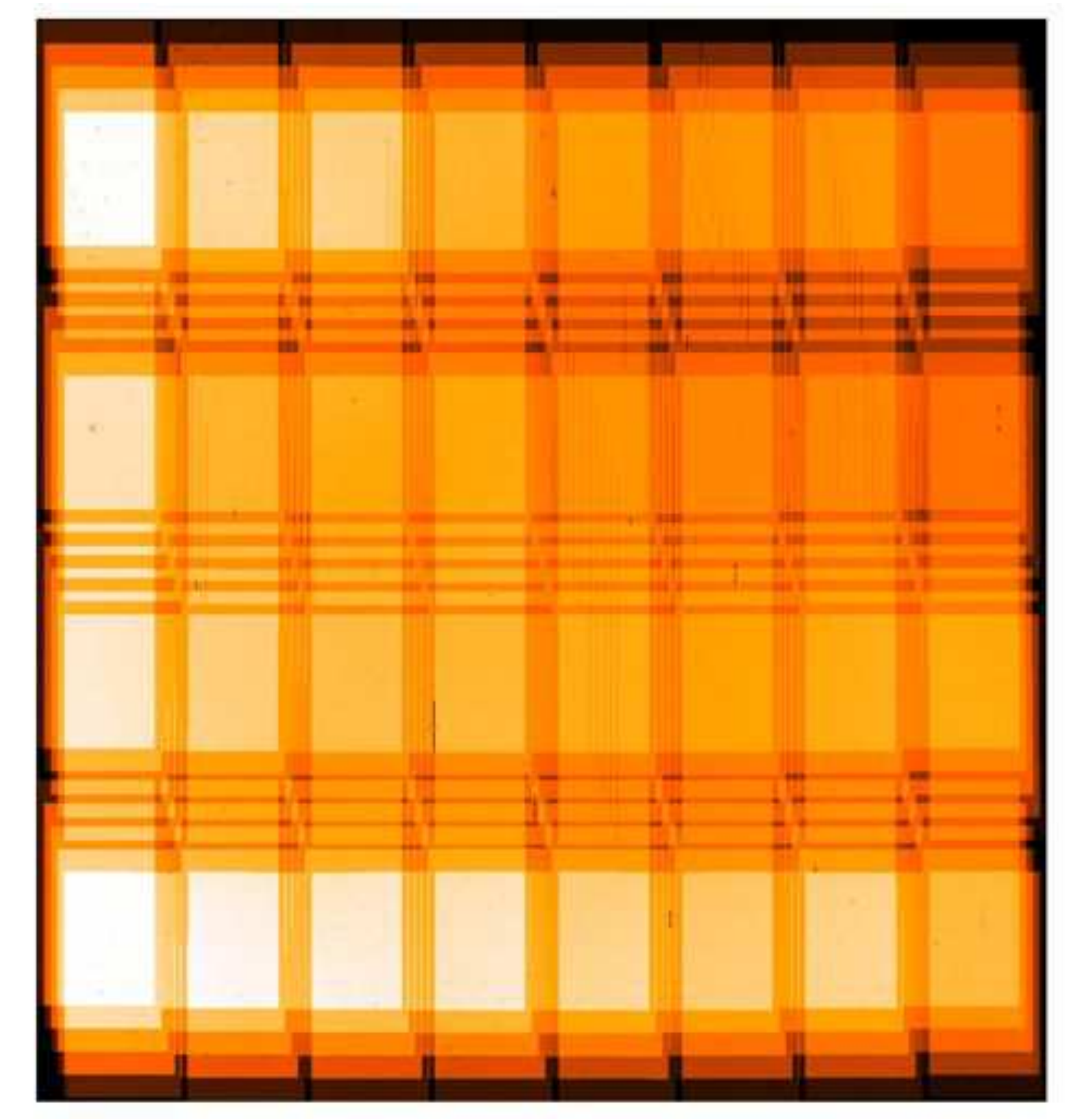}
\caption{Typical weight map for a VST mosaic ($\sim$ 1 deg$^2$) 
  obtained with the ``5-point diag'' pattern using steps of $\pm$25 
  arcsec in X and $\pm$85 arcsec in Y. The lighter regions are those 
  with higher S/N.}
\label{dithering}
\end{figure}

The colour magnitude diagram (CMD), containing stars born over the
whole lifetime of the galaxy, is a fossil record of its SFH.  
With STEP we aim at investigating the stellar
populations of the SMC with CMDs at least a couple of magnitudes
fainter than the main sequence (MS) turn-off (MSTO) of the oldest population. This is
crucial to break the age-metallicity degeneracy and safely recover the 
SFH of the investigated objects. The SFH will be derived using
the synthetic CMD technique \citep[e.g.,][]{tosi91,cignoni10}. This
kind of study has already been performed on several SMC fields both
with ground-based and HST data \citep[e.g.,][and references therein]
{dolphin01,Zaritsky2002,harris04,mccumber05,chiosi06,noel07,noel09,Weisz2013,cignoni13}. Our
plan, however, is to obtain CMDs 
significantly  fainter than the
oldest MSTO for the entire galaxy,
including its halo and the wing in the direction of the LMC. This will
allow us to infer for the first time the SFH in a homogeneous way over
the whole SMC and over the entire Hubble time.

We intend to use classical variable stars (RR Lyrae, Cepheids,
Anomalous Cepheids, $\delta$ Scuti etc.) as population tracers,
especially in the unexplored region of the Bridge. 
Indeed, these variable stars provide us insights on the
conditions at the epoch of their formation by probing 
young ages: 50--200 Myr with short and intermediate--period
classical Cepheids--CCs; intermediate-ages: $\sim$1--2 Gyr, with 
the anomalous Cepheids--ACs; and old ages: t $>$ 10 Gyr,
with the RR Lyrae stars \citep[for previous similar studies on the SMC body see
e.g.,][]{Haschke2012,Subramanian2012}. 
Eventually, the time-series images are 
stacked, allowing us to derive  deep CMDs and, in turn, to perform a
 quantitative analysis of the SFH of the investigated tiles 
(in analogy with the VISTA telescope surveys, hereinafter
we call ``tile'' one single OmegaCAM pointing),
 completing the information gathered from the analysis of the
 pulsating stars.

STEP also aims at investigating the first stages of star formation
by securing a complete mapping of pre-main sequence (PMS) objects. 
A survey of young stellar objects in the SMC was conducted with  
the Spitzer satellite by \citet{Sewilo2013} and HST
has already allowed to examine in detail the PMS in a number of SMC
regions \citep{nota06,carlson07,cignoni11}, but again only
with the wide field VST survey we will be able to infer the whole
distribution. 

STEP will allow us to identify and survey the whole population of SMC star
clusters of any type, age and mass above our detection limit. We will
make a homogeneous census of the SMC candidate clusters, and 
thus be able to ascertain whether NGC\,121 is indeed the only example
of an old system in the SMC, as currently believed, or other as old or
possibly older clusters are identifiable \citep[see e.g.,][]{Shara1998,glatt08a}.

In summary, STEP will eventually allow us to answer several open
questions: 1) Are the trends of 
SFH with position connected with the interaction history 
of the SMC? 2) Do field and
cluster components share the same SFH? 3) What are the cluster and
field age-metallicity relations ?  4) What are the evolution
properties of low metallicity stars in the mass range 1-2 M$_{\sun}$
?  5) Which is the variable star population of the Bridge?  6) Which
is the SFH of the intra-cloud population?  7) How did the stellar
component of the Bridge form? By tidal stripping or local formation
\citep{bekki07, harris07}?  8) What is the impact of metallicity on
PMS accretion and on the global properties of star formation (star
formation rate and efficiency, and initial mass function)?

There is no other completed or ongoing optical survey on nearby galaxies as
deep, wide and panchromatic as STEP. The Magellanic Clouds Photometric
Survey \citep[MCPS, see e.g.,][and references therein]{Zaritsky2000,Zaritsky2002,harris04}
provided photometry of about 16 deg$^2$  centred on the SMC. 
As we shall see in the next sections, compared to MCPS, the STEP
survey has the major advantage of observing with a spatial resolution (i.e. CCD pixel-size) 3.5 times
higher and with significantly better average seeing conditions. 
These occurrences mean that STEP can observe stars as faint
as $g\sim 22.5-23$ mag (i.e. the magnitude of the oldest TO) 
with uncertainties lower than 10-15\% even
in the most crowded regions of the SMC.
The 
 OGLE III survey \citep[see]{Udalski2008} provides an unpaired amount of information 
 for the variable stars in the SMC body, but, again, it is much shallower than
STEP and does not cover the Bridge. The next generation OGLE survey
(OGLE IV) covers a wider area around the Magellanic System, including
the Bridge, but is still shallower than STEP.  The Southern Sky Survey
(http://msowww.anu.edu.au/skymapper/survey.php) that is being performed
with the Skymapper telescope \citep{Keller2007}, will observe the SMC and the Bridge, but
at shallower magnitude limit (g$\sim$22.9 mag with
S/N\footnote{S/N=Signal to Noise}$\sim$5) than
STEP and with a worse spatial resolution. Furthermore, their
planned 6 epoch sampling is too coarse for an accurate identification and
characterisation of the Bridge variables.
First results of the Outer Limits Survey, an NOAO survey designed to
detect, map, and characterise the extended structure of the MCs, were
presented by  \citet{Saha2010}. However, contrarily to STEP,  this survey aims at
investigating only the outer regions of the MCs. 
NOAO observations to further sample the Magellanic stellar periphery
have recently began 
as part of the Survey of the MAgellanic Stellar History  \citep[][]{Olsen2014}.

\begin{figure*}
\includegraphics[width=18cm,angle=90]{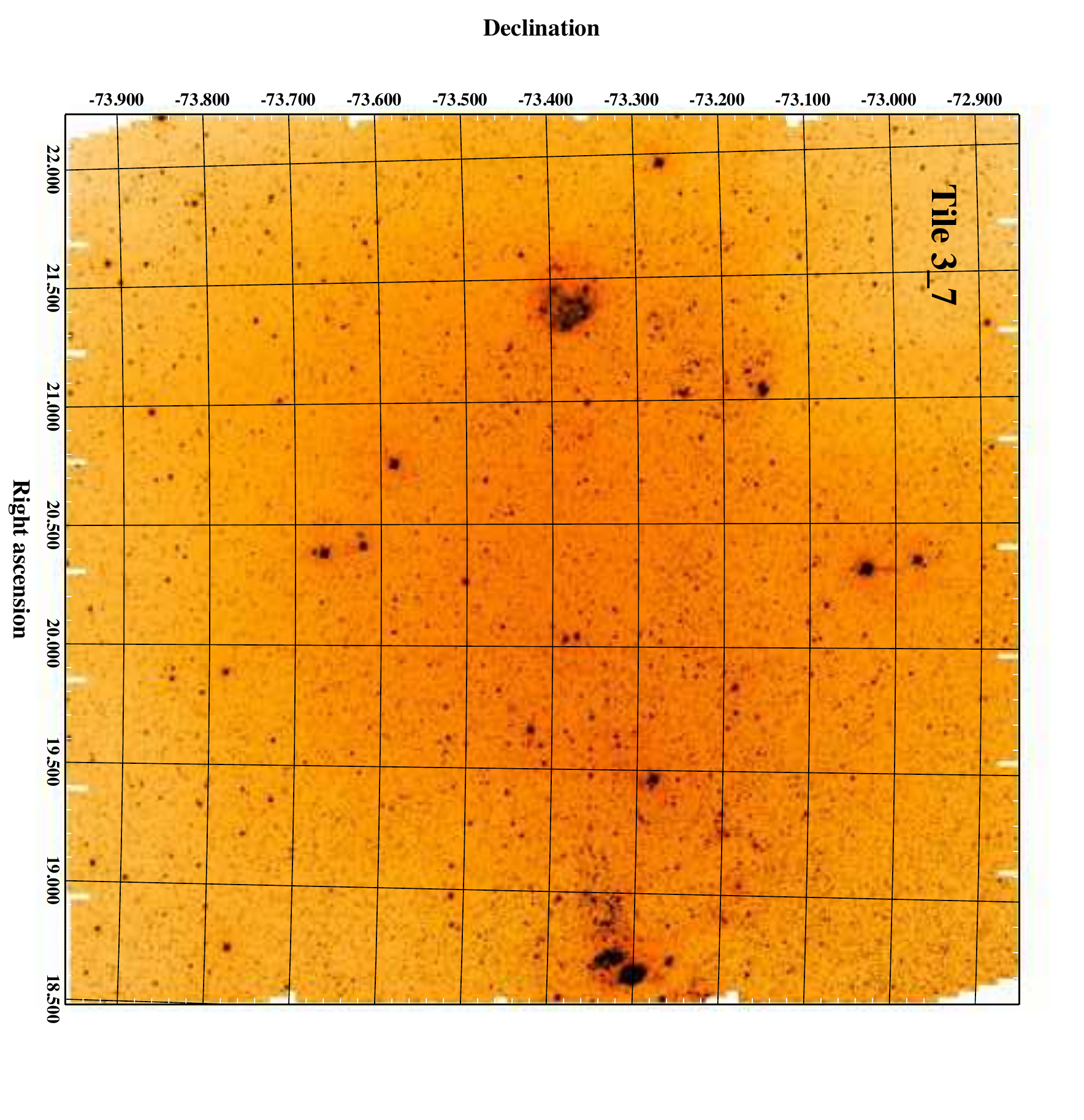}
\caption{$g$-band VST plate of tile 3\_7. North is up and East is 
  to the left.}
\label{mapAllT37}
\end{figure*}

\subsection{Observing strategy of the STEP Survey }

To address the questions listed above, we proposed, and
obtained, to use part of the VST GTO time allocated by ESO to 
Istituto Nazionale di Astrofisica (INAF) in return for the procurement of the telescope and 
acquire $g$, $r$, $i$, $H_{\alpha}$\footnote{See
  Sect~\ref{vstTelescope} for a discussion about the VST filter
  system}
photometry for 72 squared degrees covering the whole SMC body,
the Bridge and 2 deg$^2$ of the Magellanic Stream down to a limiting
magnitude (AB system) $g\sim$24 mag with S/N=10, and $r~(H_{\alpha})\sim22.5$ mag
with S/N=5. The survey is organised with tiles of 1 deg$^2$ each,
partially overlapping with each other to allow a homogeneous calibration.

In addition, we acquired 24-epoch time-series photometry of 8 deg$^2$
on the Bridge down to $V\simeq 19.5$ mag (i.e. reaching fainter than the
mean magnitude of the RR Lyrae stars), with S/N=100. When summed up, these images
will allow us to reach $g\sim$24 mag with S/N=10.
Originally, we planned to image with time-series the whole
Bridge. However, after the first observations in Period 88 (see
below), we found that the huge overheads in pointing and filter
changing made the observing efficiency (shutter time/total duration of
the exposure) so low that we decided to
complete only the 8 Bridge fields already started and cover the remaining
Bridge fields using the same observing strategy adopted for the SMC
body,  (i.e., without time-series).

Details on the instrumentation and the observing strategy are
reported in the following subsections, while the data
reduction is discussed in Sect. 3.

\begin{figure*}
\includegraphics[width=18cm,angle=90]{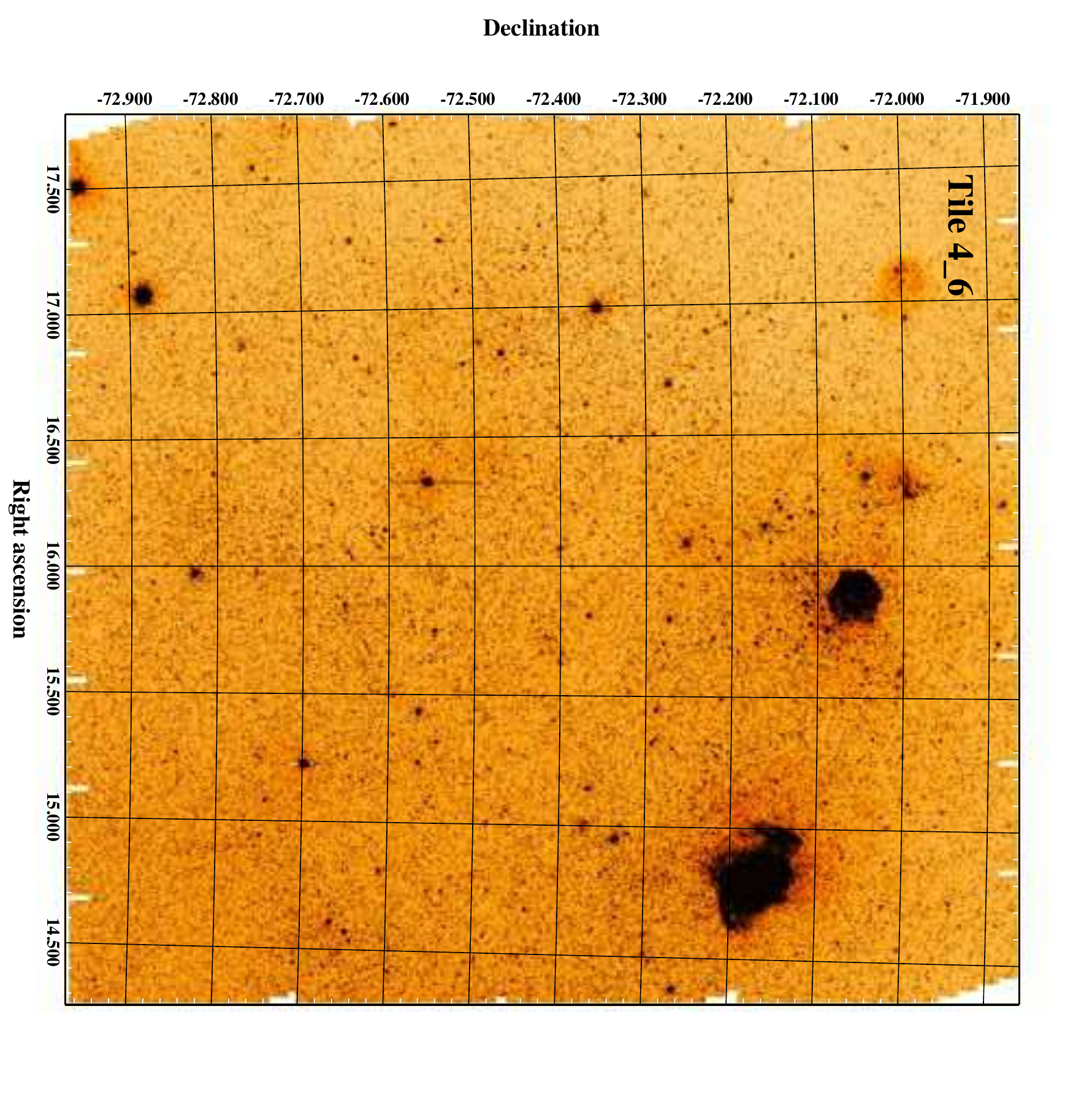}
\caption{$g$-band VST plate  of tile 4\_6. North is up, East is to 
  the left. }
\label{mapAllT46}
\end{figure*}

\subsection{The VST telescope and OmegaCAM}
\label{vstTelescope}

\begin{figure*}
\includegraphics[width=18cm]{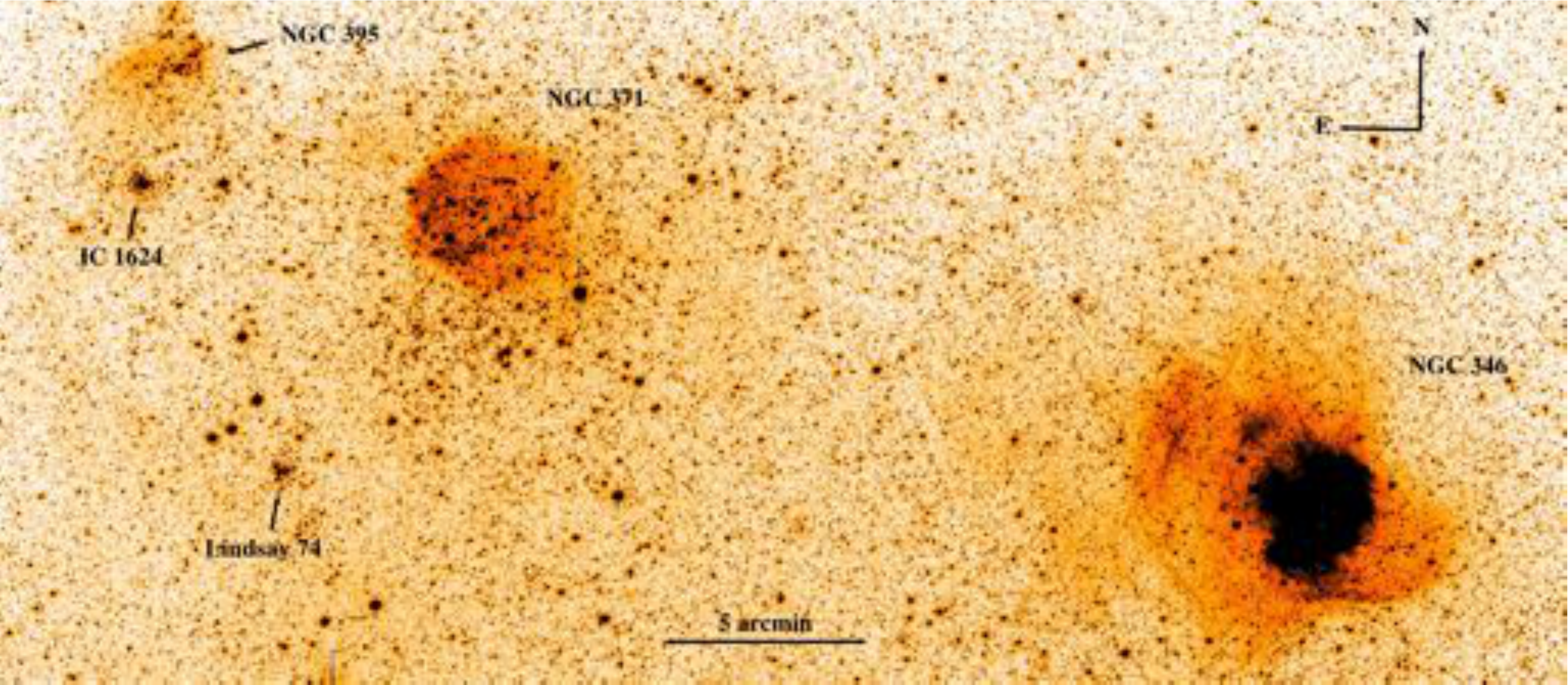}
\caption{Enlargement of Fig. ~\ref{mapAllT46} showing the northern part of tile 4\_6, including the 
 well known star forming region NGC\, 346 and several other interesting 
 clusters and associations which are labelled in the figure.}
\label{ngc346a}
\end{figure*}
 
The VST (built by the INAF-Osservatorio
Astronomico di Capodimonte, Naples, Italy) is a 2.6-m wide field
optical survey telescope \citep{VST}. 
VST is placed at Cerro
Paranal, Chile on the same platform as the VLT 8-m telescopes.  
VST has a f/5.5 modified Ritchey-Chretien optical layout on an alt-azimuth
mounting. It features a two lens wide-field corrector, with the dewar
window acting as a third lens (and an optional atmospheric dispersion
compensator - ADC) giving a correct FoV of 1 deg$^2$ 
\citep[see][]{Schipani2012}. 
The primary mirror is a concave 2.6 m hyperbolic
meniscus axially supported by 84 active supports, placed on four
rings and laterally sustained by passive astatic levers. 
The convex hyperbolic secondary mirror is controlled by  a
hexapod with 5 degrees of freedom. 
The telescope is equipped with OmegaCAM, a 1 deg$^2$ camera build by a
consortium of european institutes \citep{Kuijken2011}.
The camera is a 32-CCD, 16k x 16k detector mosaic with 
0.214 arcsec per pixel scale (for the two-lens corrector, 0.215 for
the ADC configuration). The CCDs are thinned, blue-sensitive, 3-edge
buttable CCD44-82 devices of high cosmetic quality made by e2v. The field
distortion is very low, and the image scale is practically constant
over the whole field of view. The gaps between the CCDs are rather
narrow, and the overall geometric filling factor of the array is
91.4\%.

OmegaCAM can handle 12 filters that can be changed during the night. 
 Currently, the available filters include the Sloan Digital Sky Survey
 (SDSS) $ugriz$ set, Johnson $B$ and $V$
filters, plus several narrow-band filter mosaics such as
$H_{\alpha}$. It is important to note that the VST filter system is actually based on
the USNO $u^{\prime}g^{\prime}r^{\prime}i^{\prime}$ system \citep{smith02} which is
slightly different from SDSS $ugriz$ \citep[for details, see ][]{Fukugita1996,Gunn}. 
This is shown in Fig.~\ref{figureFilter} where the
throughput of the filters used in the present survey are displayed and
compared with those of the SDSS (properly rescaled to
improve the comparison). 

However, since we will calibrate our data on the
``natural''  SDSS system,  in this paper we use the $ugriz$ notation.
The OmegaCAM calibration plan  ensures that all data
can be photometrically and astrometrically calibrated to 0.05
mag and 0.1 arcsec rms precision, respectively.

\subsection{STEP area coverage}

The main aim of the survey is to image the body of the SMC and the
Bridge. The choice of the survey fields was driven by:  

\begin{itemize}
\item
Complete coverage of the SMC+Bridge area. In particular, we tried to include
the classical SMC diameter limit at $B\approx25$ mag arcsec$^{-2}$ \citep{bot88}
as well as major features traced by the distribution of stars, star clusters
\citep[e.g.,][]{irw91,bic08} and H\,{\footnotesize I\normalsize}
gas \citep[e.g.,][]{sta03,hat05,mul03}, for both the SMC and the Bridge.

\item
Maximize the overlap with the VMC survey \citep[see][]{Cioni11}.

\end{itemize}

In order to accomplish these tasks efficiently we used the Survey Area
Definition Tool \citep[SADT]{arn08}. A geodesic rectangle with
vertexes $\alpha$=23$:$53$:$54; 04$:$29$:$30; $\delta=-$75$:$54$:$46;
$-$70$:$02$:$00 was generated as the basis of the tile creation
process.  The unnecessary tiles were removed leading to the coverage
shown in Fig.~\ref{mapStep}. This figure shows the 72 tiles (coloured
boxes), which define the area surveyed by STEP (apart the two tiles
placed into the Magellanic Stream). The tiled area covers almost 
completely the distribution of stellar clusters and associations (black
dots) and maximizes the overlap with the VMC survey (grey boxes).

The coordinates of the centres of each STEP tile are given in Table~\ref{tabFields}. 
Each tile is identified by two numbers: the first number
gives the row and the second the column of the position of the
tile in the mosaic that covers the surveyed area. 
Row numbers increase from South to North and
column numbers increase from West to East.

In the process of defining the mosaic, SADT requires as input the
observing parameters that are associated to small (i.e. jittering) and
large (i.e. mosaicking) displacements in the tile position. For the
STEP survey the maximum jitter was set to $10^{\prime\prime}$, and the
tile overlap in $\alpha$ and $\delta$ to $120^{\prime\prime}$.

\subsection{STEP observations}
\label{sectStepStrategy}

We can distinguish our observations in ``deep'' (on the SMC body, on the
Wing and on part of the Bridge) and ``time-series'' (seven fields in
the Bridge and one in the SMC periphery).
Table~\ref{observingStrategy} describes the main parameters of the VST
observations, including the exposure times.  For the
time-series we decided to take a single exposure per epoch and
filter. The images at different epochs of each tile were slightly dithered  to
cover the gaps between CCDs in the stacked images.  As for the deep exposures, we
divided the total exposure time, as estimated by the OmegaCAM
Exposure Time Calculator
(ETC)\footnote{http://www.eso.org/observing/etc/bin/gen/form?INS.NAME\\=OMEGACAM+INS.MODE=imaging}
into 5 or 10 sub-exposures (see Table~\ref{observingStrategy}).  To
avoid saturation for the bright SMC/Bridge stars we also acquired 5
short (25 s each) exposures per tile.  We chose the same dithering pattern
for both time-series and deep exposures. In particular, we adopted the
``5-point diag'' pattern\footnote{See the omegaCam template manual,
  document available at http://www.eso.org/sci/facilities/paranal/instruments/omegacam\\/doc/VST-MAN-OCM-23100-3111-2\_7\_1.pdf}
which offsets the telescope by the same amount between
successive exposures. The size of the steps in X and Y directions is
as large as the largest gap in the mosaic, i.e.  about $\pm$25 arcsec
in X and $\pm$85 arcsec in Y.  A typical weight map for the stacked
tile  is shown in Fig.~\ref{dithering}, where the brighter 
 regions are those with larger S/N. Although the  weight map
is provided as output by the VST--Tube package, we
did not use it in the derivation of the photometry (see next sections).

For the deep images we required clear sky, but we allowed for
thin--cirrus for the time--series photometry, to increase the
probability of program execution.
Moreover, images obtained in each of the two filters were
left free to be acquired in different nights. Hence, to
perform the photometric calibration of the data we need secondary
standard stars. 
For time-series tiles this was secured  by requiring that at
least two of the epochs were taken under photometric condition. 
To calibrate the deep tiles, instead, we planned for a  
couple of short exposure $g$, $i$ images (see Table~\ref{observingStrategy}) to be acquired
during a photometric night. These images were sufficiently deep to 
provide a huge amount of stars to be used as secondary standards.

Finally, we exploited the overlap between adjacent tiles to correct for 
residuals in the photometric zero points. 
The overall zero point of the photometry was checked using
calibrations obtained during different observing runs.

\begin{figure}
 \centering 
 \includegraphics[width=8.5cm]{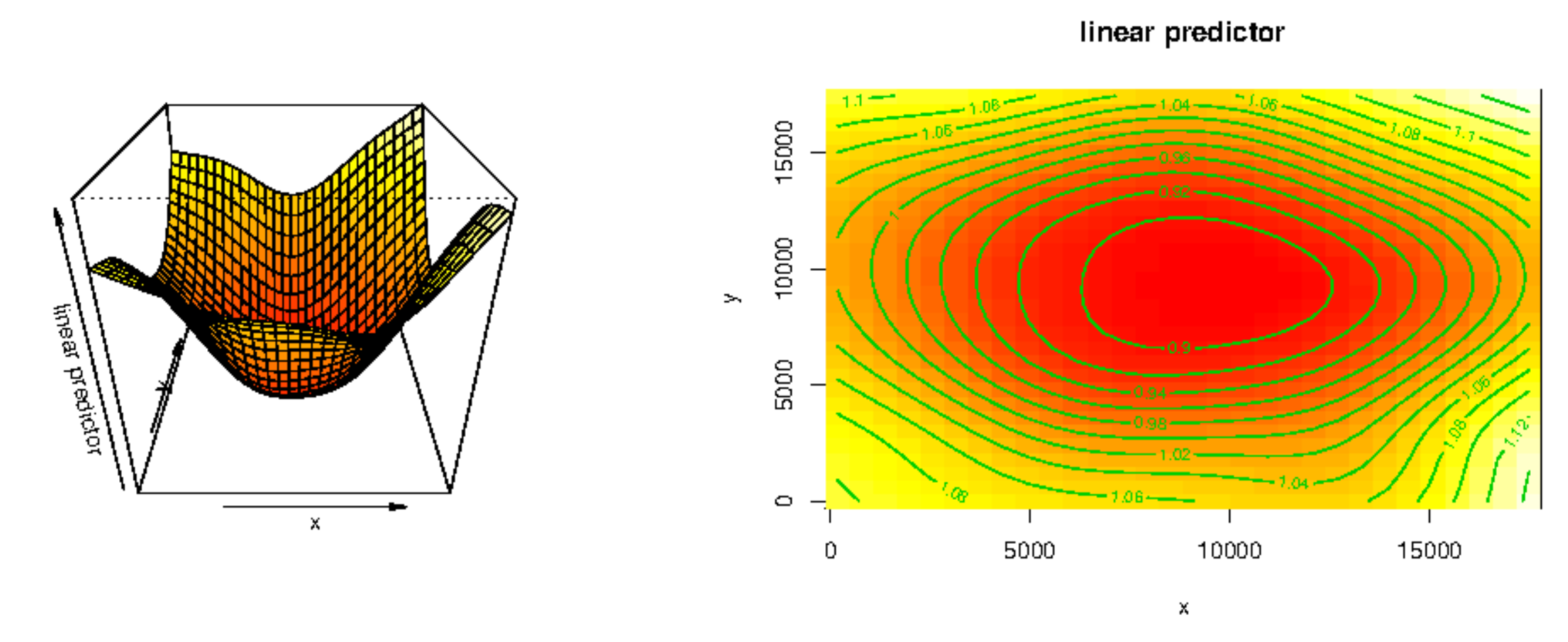}
 \caption{ Left: IC 3D view; Right: Contour plot of the 
   IC image}
 \label{fig:IC1}
\end{figure}

\begin{figure}
 \centering 
 \includegraphics[width=8.5cm]{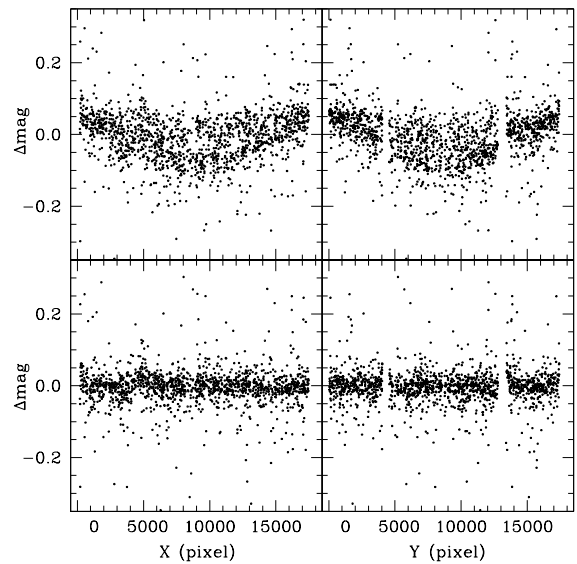}
 \caption{Residuals (VST-SDSS) vs x and y position before (top panels)  
and after (bottom panels) the IC correction }
 \label{fig:IC2}
\end{figure}

\begin{table}
\caption{Observing Strategy}
\label{observingStrategy}
\begin{center}
\begin{tabular}{ccc}
\hline
\hline 
\noalign{\smallskip} 
Period & T$_{\rm exp}$($g$) & T$_{\rm exp}$($i$)  \\
\noalign{\smallskip}
\hline
\noalign{\smallskip} 
88-90 &  5$\times$25 s; 5$\times$520 s & 5$\times$25 s; 5$\times$520 s\\
91-92 &  5$\times$25 s; 10$\times$300 s & 5$\times$25 s; 10$\times$300 s\\
\noalign{\smallskip}
\hline
\noalign{\smallskip}
\multicolumn{3}{c}{Photometric calibration}\\
\noalign{\smallskip}
\hline
\noalign{\smallskip}
88-92 &   1$\times$45 s & 1$\times$45 s \\
\noalign{\smallskip}
\hline
\noalign{\smallskip}
\multicolumn{3}{c}{Time-Series}\\
\noalign{\smallskip}
\hline
\noalign{\smallskip}
88-91 &   1$\times$25 s; 1$\times$120 s & 1$\times$25 s; 1$\times$180 s \\
\noalign{\smallskip}
\hline
\noalign{\smallskip}
\end{tabular}
\end{center}
\end{table}

Most of the observations of the STEP Survey are obtained in service mode
by ESO staff. This guarantees efficiency of operations and a high
level of data homogeneity. Only a few hours of observation were conducted in
visitor mode (GTO compensation time).

The constraints on the observing conditions varied as a function of
the crowding of the target regions. In general, for the SMC body and the
Wing we required a better seeing than for the less crowded tiles in
the Bridge. A summary of the observing constraints is provided in 
Table~\ref{observingConstraints}.

Since the SMC and the Bridge never rise above $50^\circ$ from the
horizon, the seeing measured on the images is expected to be 
significantly larger than that measured at the Zenith. Hence to keep a
reasonable probability of realisation of the observations, we decided
to  accept a seeing limit of 1.0$^{\prime\prime}$--1.1$^{\prime\prime}$
(corresponding to $\sim$0.7$^{\prime\prime}$ at the Zenith)
even for the crowded body of the SMC. These values of seeing allow us to
reach our scientific goal in terms of magnitude limit.

\begin{figure*}
\hbox{
\includegraphics[width=9cm]{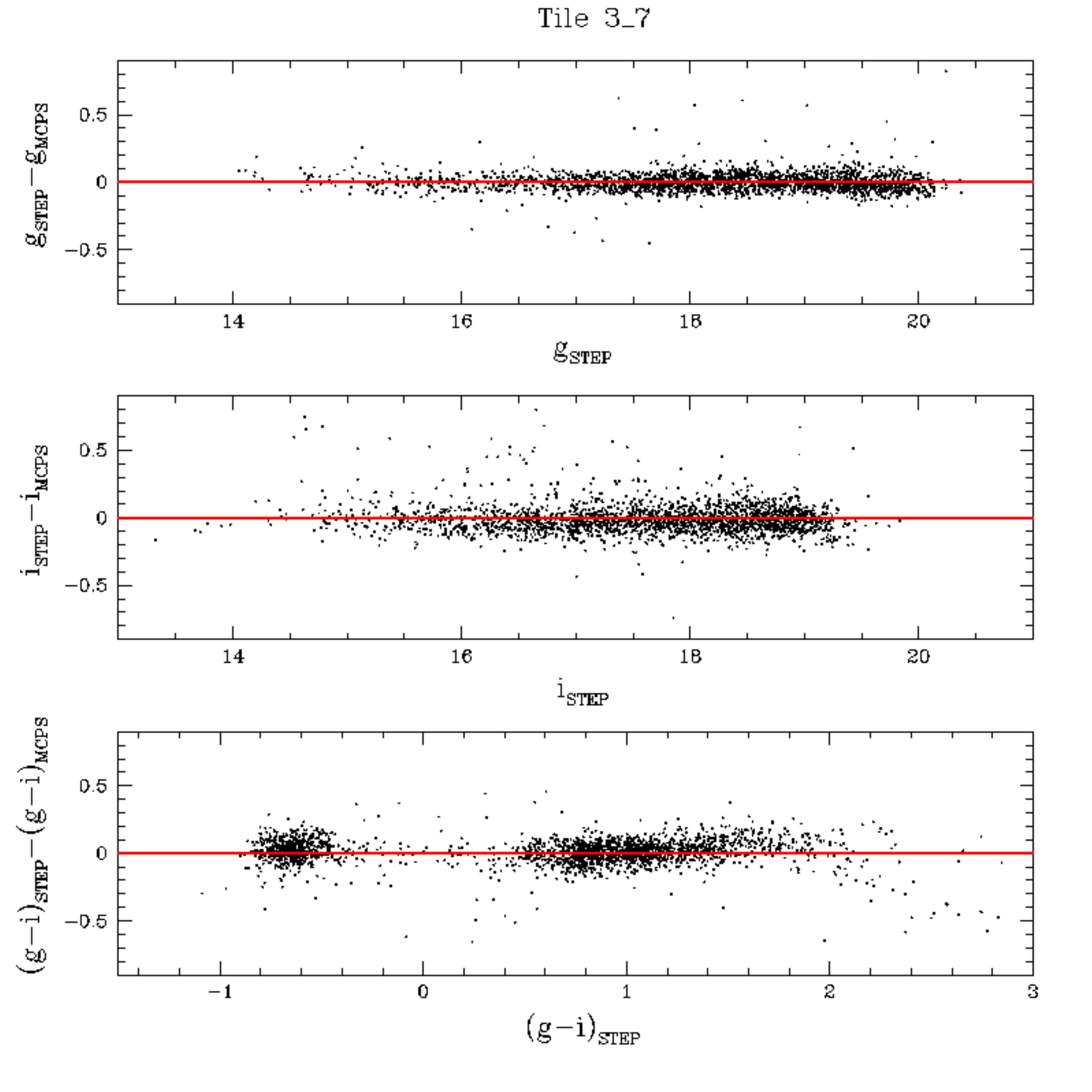}
\includegraphics[width=9cm]{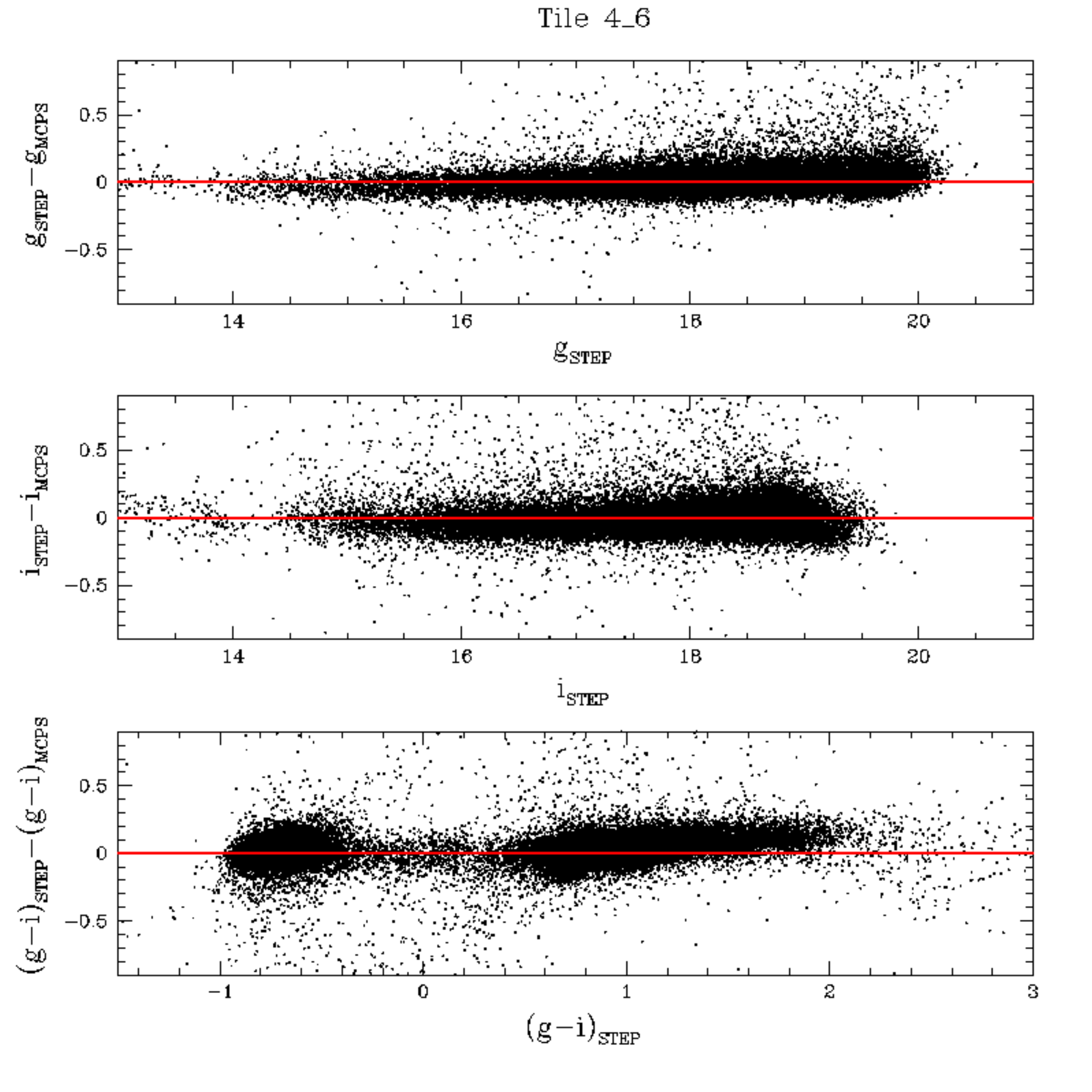}
}
\caption{Comparison between STEP and MCPS  photometries 
  \citep[transformed to the SDSS system according to][colour 
  conversions]{Jordi2006}.}
\label{comparison}
\end{figure*}

\begin{figure*}
\hbox{
\includegraphics[height=8.5cm]{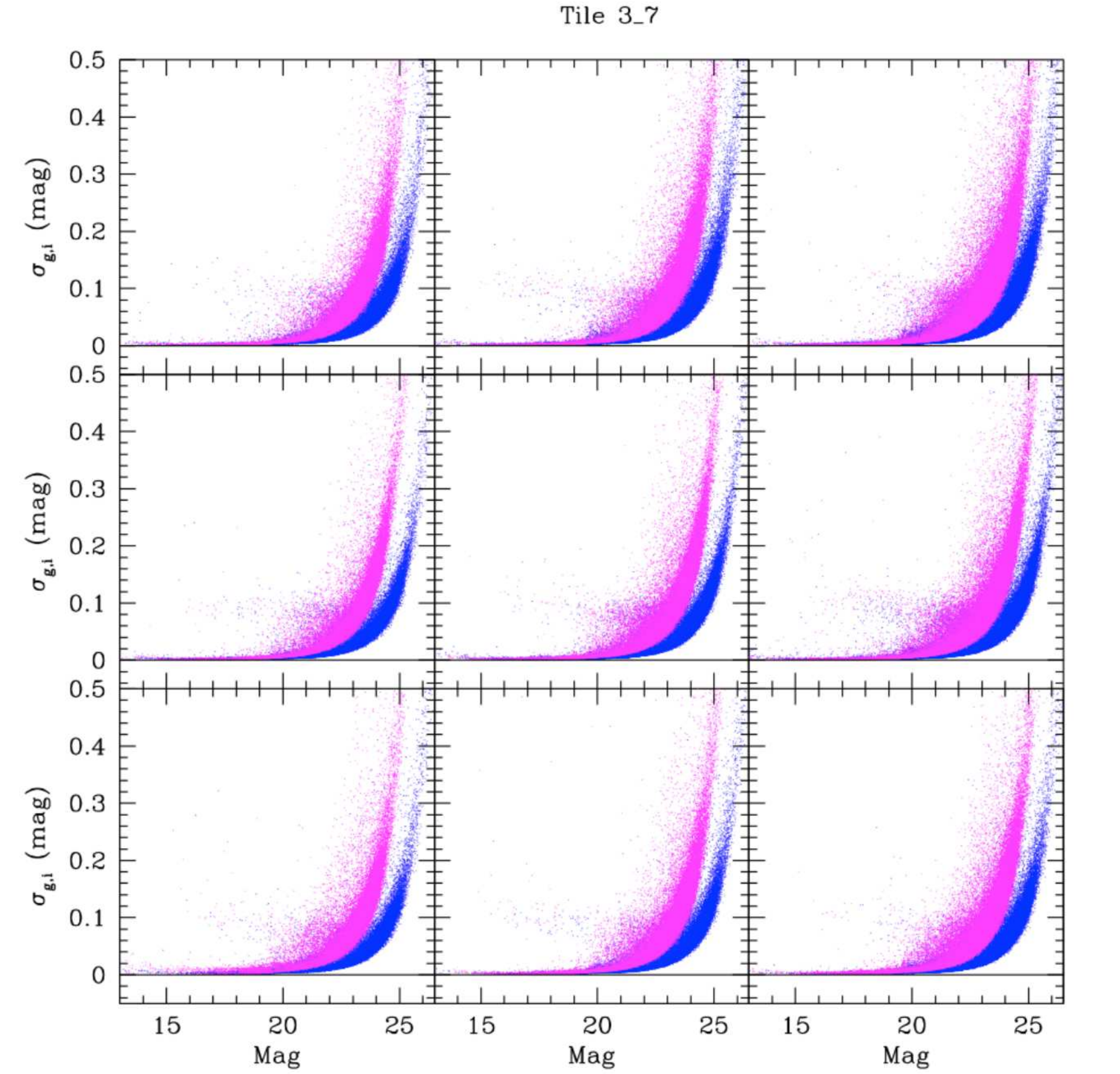}
\includegraphics[height=8.6cm]{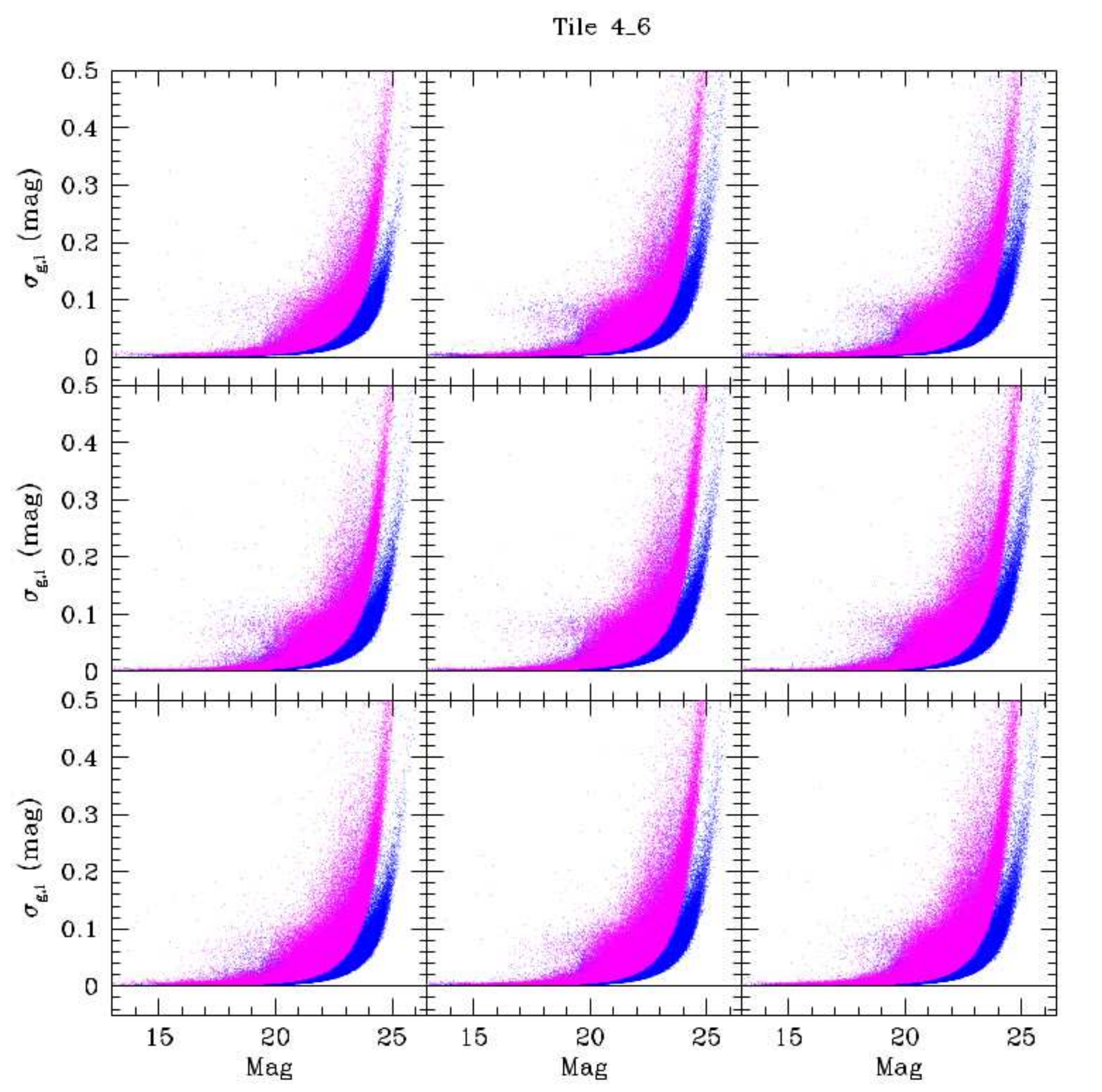}
}
\caption{Photometric errors in $g$ (blue) and $i$ (magenta) for  tiles 3\_7 and 4\_6. To verify the presence of 
possible differences  within each image, the tiles were divided into 
nine sub-frames 20$^{\prime}$$\times$$21^{\prime}$ each. North is up 
and east on the left}
\label{rms}
\end{figure*}

A more relaxed constraint of 1.4$^{\prime\prime}$ seeing was used for the
uncrowded regions of the Bridge to be imaged in time-series mode. 
For the Bridge tiles to be observed with one epoch deep images, this constraint
was reduced to 1.1$^{\prime\prime}$ to account 
for the shorter total exposure time with respect to those obtained by summing
up all the time-series images. 

\begin{table}
\caption{Observing constraints. Seeing is to be intended
as the full width at half maximum measured on the image. TS stands for time-series.}
\label{observingConstraints}
\begin{center}
\begin{tabular}{ccccc}
\hline
\hline
\noalign{\smallskip} 
Field & Seeing & Moon & Airmass & Weather \\
\noalign{\smallskip}
\hline
\noalign{\smallskip} 
SMC  &  1.0$^{\prime\prime}$--1.1$^{\prime\prime}$ & 0.5   &  1.8  & Clear\\
Bridge  & 1.1$^{\prime\prime}$ & 0.5   &  1.8  & Clear\\
Bridge (TS) &  1.4$^{\prime\prime}$ & 0.8   &  1.8  & Thin cirrus\\
Phot.  Cal.  &  1.5$^{\prime\prime}$ & 0.8   &  1.8  & Photometric\\
\noalign{\smallskip}
\hline
\end{tabular}
\end{center}
\end{table}

\begin{table}
\caption{Distribution of the observations in the  ESO Periods
  (column 1). Columns (2) and (3) show the hours allocated and those actually devoted to
the observations, respectively. Column (4) list the efficiency
(in percentage) of the observations, i.e.  the ratio
col. (3)/col. (2). Column (5) shows the percentage
of OBs graded by ESO as A,B,D. See text for a detailed explanation.}
\label{esoPeriods}
\begin{center}
\begin{tabular}{ccccc}
\hline
\hline
\noalign{\smallskip}  
Period & Allocated & Observed & Obs./Allocated &Grade A,B,D \\
           &   (h) &   (h) & (\%)  &(\%)\\
 (1) & (2) & (3) & (4) & (5)  \\
\noalign{\smallskip}
\hline
\noalign{\smallskip}   
88  & 40 & 9.8  & 24.5  &16,30,54  \\
89  & 43 &37.2 & 86.5  &19,37,44 \\
90 &  34 &24.8 &  73 &29,65,6 \\
91 &  49 &19.7  & 40 &20,44,36 \\
92 &   16 &  7.2   & 45 &50,32,18  \\
93 &   18  &       &       &                \\
\noalign{\smallskip}  
\hline
\noalign{\smallskip} 
       & Total & Total  & Average & Average\\
\noalign{\smallskip}
\hline
 88-92   &   182 & 98.7& 54  & 27,42,31 \\
\noalign{\smallskip}  
\noalign{\smallskip}
\hline
\end{tabular}
\end{center}
\end{table}

The efficiency (hours of actual observations/hours allocated) of the
STEP observations is shown in Table~\ref{esoPeriods}.
A quick comparison of columns 2 and 3 reveals that, at face value, the observing
efficiency was of the order of 54\%. However it can be noticed that the
actual efficiency was significantly lower. 
Table~\ref{esoPeriods} also reports the average grade\footnote{For the
  meaning of the various grades see
  http://casu.ast.cam.ac.uk/surveys-projects/vista/data-processing/eso-grades} 
from Period 88
to 91. It can be seen that grades ``A'' and ``B''  were obtained only in
27\% and 42\% of the cases, respectively.  The remaining 31\% of the
images was graded as ``D'', i.e. with 
characteristics that are by more than 20\% outside the limits we had set, in
terms of seeing and/or ellipticity. This last occurrence
is particularly harmful because it prevents us from obtaining a good
astrometric solution for the images composing one mosaic, making the
tiles sometimes completely useless, and to be re-observed in  the
following semester. 
Hence, even if the observing  efficiency of the STEP observations was
formally of
54\%, the survey efficiency, i.e. the percentage of useful data, was
well below 50\%.

\begin{figure*}
\hbox{\includegraphics[width=9cm]{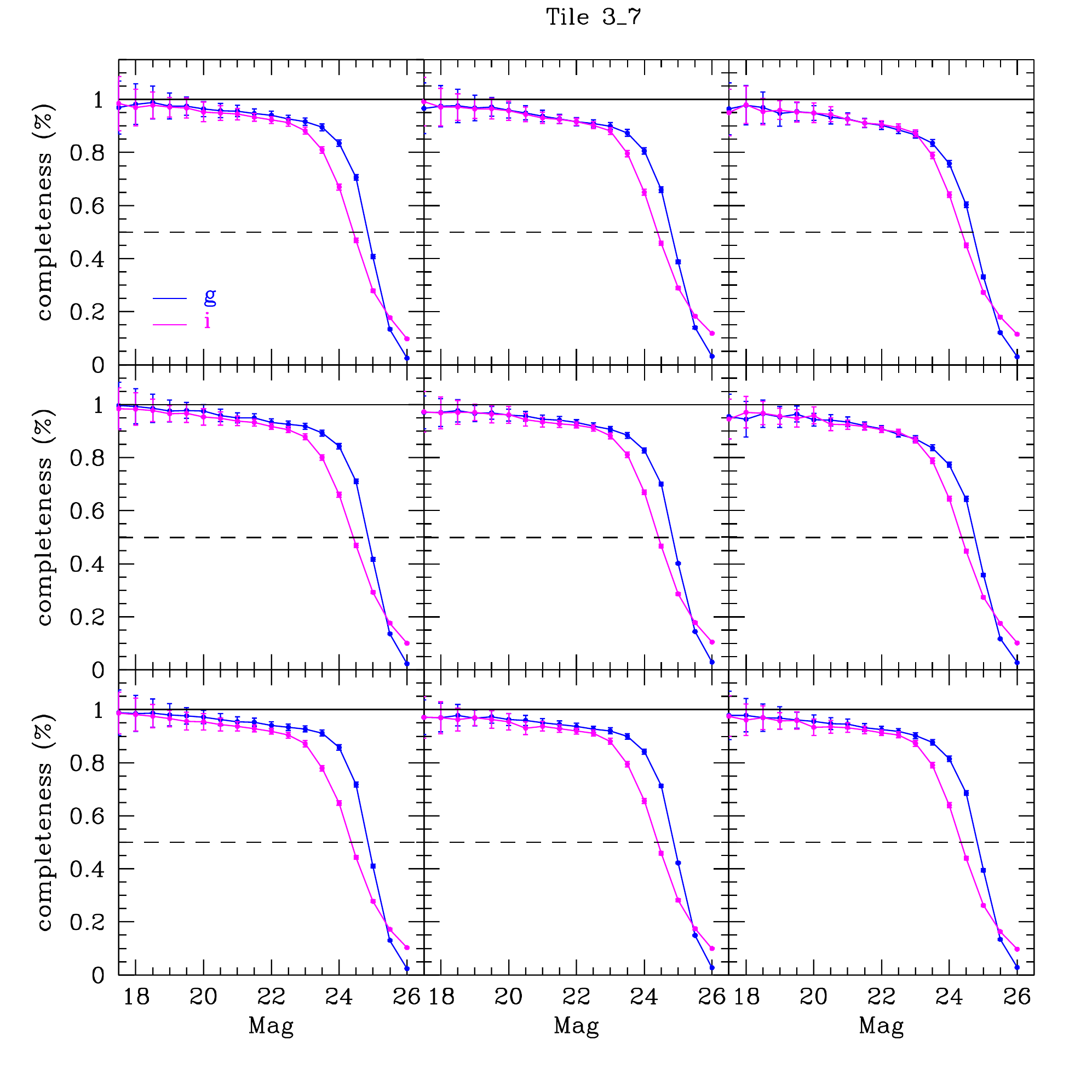}
  \includegraphics[width=9cm]{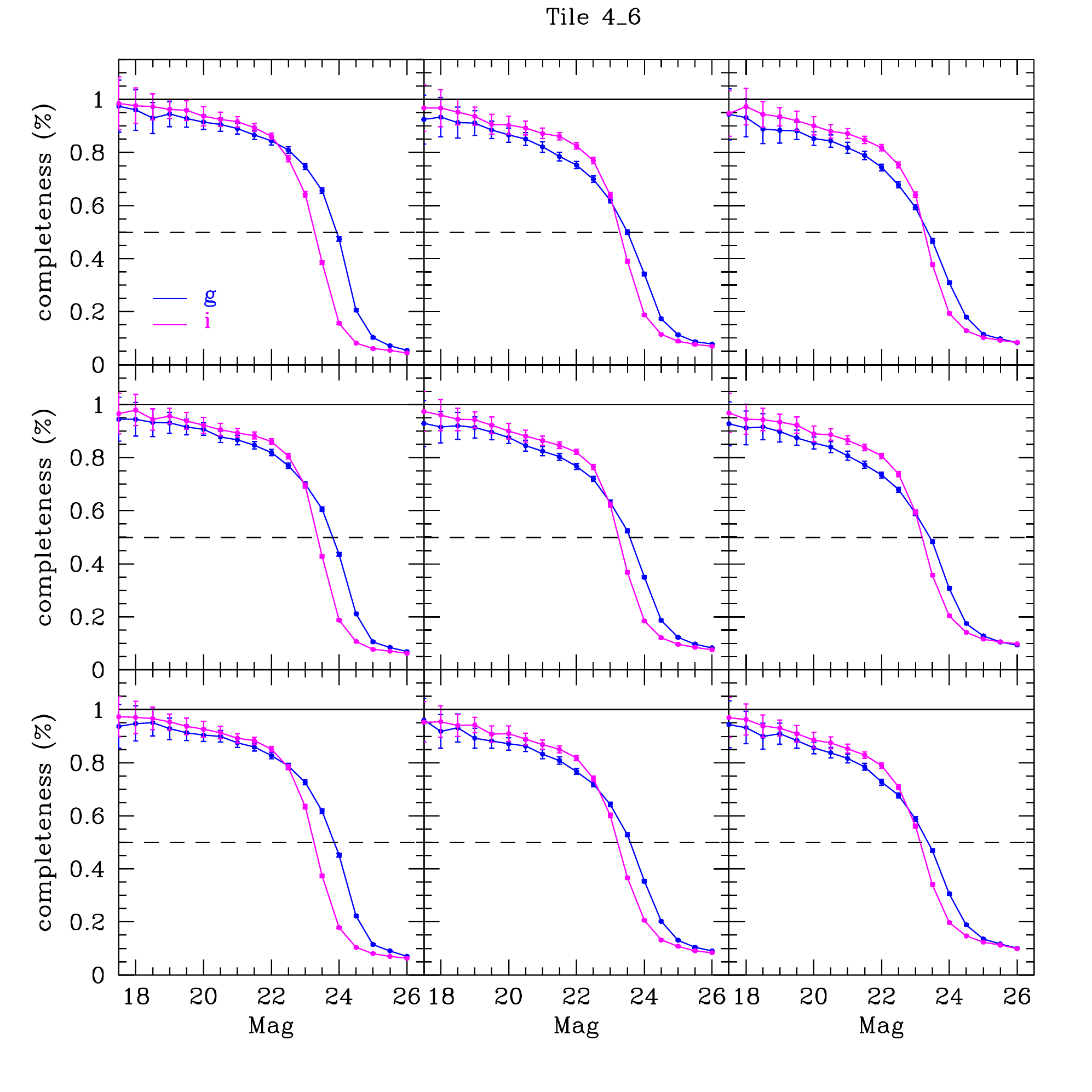}}
\caption{Completeness for the tiles 3\_7 and 4\_6 (blue and magenta 
solid lines refer to  $g$ and $i$ bands, respectively). To verify the presence of 
possible differences in completeness within each image, the tiles were divided into 
nine sub-frames 20$^{\prime}$$\times$$21^{\prime}$ each. North is up 
and east to the left. The 100\% and 50\% levels of completeness are 
shown with black solid and dashed lines, respectively.}
\label{completeness}
\end{figure*}

\subsection{STEP progress} 

As shown in Table~\ref{esoPeriods}, STEP observations started on
P88. Time-series photometry was collected till P90. The
number of epochs for each tile is listed in Tab.~\ref{epochs}. Given
the extremely low efficiency (shutter time/total duration of OB) 
of this kind of observations (as low as 25\%) we decided to acquire 
only deep tiles starting from P91. 

As for the deep tiles, at the moment of writing, there are 16 tiles
completed (see Table~\ref{tabFields} and Fig.~\ref{mapStep}), whereas
several additional tiles were so far only partially observed (see
Table~\ref{tabFields}).  The observations in $r, H_{\alpha}$ will
start as soon as the $g$, $i$ survey is completed.

\begin{table}
\caption{Number of epochs for each tile observed by means of
 time-series photometry.}
\label{epochs}
\begin{center}
\begin{tabular}{ccc}
\hline
\hline
\noalign{\smallskip}  
Tiles & n$_{\rm epochs}$ ($g$) & n$_{\rm Epochs}$ ($i$) \\
\noalign{\smallskip}
\hline
\noalign{\smallskip}   
5\_9   &  26   &      23      \\               
2\_10 &  23    &     22      \\
3\_12 &  24    &     23       \\
3\_13 &  24    &     23        \\
3\_14 &  23    &     22        \\
3\_15 &  21    &     20        \\
3\_17 &  22     &   21         \\
3\_19 &  21     &   20         \\
\noalign{\smallskip}  
\hline
\end{tabular}
\end{center}
\end{table}

\subsubsection{The first two completed tiles.}

We have chosen to give highest priority to observations of 
two tiles which include fields already observed by our group with the
HST Advanced Camera for Surveys (ACS): one in the most active region
of the SMC main body and the other in the most external regions of the
Wing. The former is tile 4\_6, around the very young cluster NGC\,346,
and the latter is tile 3\_7 in the Wing. They cover the NGC\,346 and
the SFH9 ACS fields observed respectively by the HST programs 10248
\citep[PI Nota, see][]{nota06} and 10396 \citep[PI Gallagher,
see][]{sabbi09}.  This allows an immediate comparison of the VST
photometry with the exquisite one from ACS, as well as an independent check on the
calibration.
The plates shown in Figures~\ref{mapAllT37} and~\ref{mapAllT46} report
the OmegaCAM@VST mosaic images relative to tiles 3\_7 and 4\_6,
respectively.  Figure~\ref{ngc346a} exhibits an enlargement of the
northern part of plate ~\ref{mapAllT46}, including the
  well known star forming region NGC\, 346 and several other interesting
clusters and associations. 

Since STEP covers the whole SMC, it will eventually include not only
all our HST fields, but also those studied and published in the
astronomical literature,
both from space and from the ground.

\section{Data reduction}

In this section we describe the procedure adopted for the treatment of
the raw data and the production of the final catalogue ready for the scientific exploitation.
In the following we refer mainly to the deep exposures. Details on the
treatment of time-series data will be provided in a dedicated paper.

\subsection{Pre-reduction, astrometry and photometric calibration}
\label{prereduction}
The images of tiles 3\_7 and 4\_6 were collected during different
runs (see Table~\ref{ObsLogstep} for details). The data reduction has made use of the
VST--Tube imaging pipeline \citep{grado}, installed and running on the computers
of the INAF--VST Center
(VSTCeN\footnote{http://vstportal.oacn.inaf.it/} hosted by the INAF-Osservatorio Astronomico
di Capodimonte, Naples).
This pipeline was specifically developed for
the VST telescope but is available also for
other existing or future single or multi-CCD cameras.  Removal of
instrumental signatures that include overscan, bias and flat-field
correction, CCD gain equalisation and illumination correction have
been applied. Relative and absolute astrometric and photometric
calibration were applied before stacking for the final co-added image.
In detail, the overscan correction uses the median value of the 
overscan region portion and  the master-bias comes from a sigma-clipped
average of bias frames.  The gain variation over the field of view is
obtained using a suitable combination of twilight flat-fields and
science images (with large enough dithering) as given by the formula:

\begin{equation}
MFlat =
 Imsurfit \left(\frac{SFlat}{TFlat} <TFlat>
 \right)  
TFlat \times {\rm IC} 
\label{masterflat}
\end{equation} 

\noindent
where $MFlat$, $SFlat$, $TFlat$ stands for Master-Flat, Sky-Flat and
twilight-flat, respectively.  $Imsurfit$ is a Chebyshev polynomial
image fit, whereas IC (Illumination Correction) is an analytic function reproducing the image
concentration effect.  TFlat images are first corrected by overscan
and then corrected by the master-bias.  The average Tflat is
produced using a sigma clipped prescription.  $<$ TFlat $>$ is a median
computed on an inner region of the TFlat. The Sky-Flat is
produced with the same procedure as the TFlat but using
science images.  The Sky-Flat is used to correct the large scale
variations due to non-uniform illumination and the high S/N
TFlat is used to correct the high-frequency pixel to pixel
sensitivity variations.

To have the same zero-point for all the mosaic chips, a gain
equalisation procedure has been used. The procedure finds the relative
gain correction which gives the same background level in adjacent
CCDs.  For the OmegaCAM@VST images it was necessary to apply a further
correction caused by scattered light. This is frequent for wide field
imagers where telescope and instrument baffling can be an issue. This
results in a centrally localised additive component to the background,
and the flat field does not accurately estimate of the spatial
response of the detector.  Indeed, if not corrected, after
flat-fielding the image background would appear perfectly flat but the
photometric response would be position dependent \citep{IC}.  This
error can be mitigated through the determination and application of
the IC map. The IC map was derived using SDSS DR8 stars in properly
selected fields. The magnitude residuals (VST-SDSS) as function of the
position were fitted using a generalised adaptive method (GAM) in
order to obtain an IC map used to correct the science images during
the pre--reduction stage. The GAM allows to obtain a well behaved
surface also in case the field of view is not uniformly sampled by 
standard stars. As an example we show in Figures~\ref{fig:IC1}
and~\ref{fig:IC2} the IC map and the position dependency of the zero-point before and
after the IC application.

The absolute photometric calibration was computed on the
nights reported in Table~\ref{zp} comparing the observed magnitude of
standard stars with SDSS photometry. 

In Tab.~\ref{zp} we report the zero-points and colour terms obtained
using the Photcal tool\footnote{http://www.na.astro.it/~radovich/}. Since
the photometric standard star fields were observed with a span in
airmass too small for a suitable fit, an average value for the
extinction coefficient was adopted.  Relative photometric correction
among the exposures is obtained by minimising the quadratic sum of the
differences in magnitude between overlapping detections. The tool used
for such task is SCAMP \citep{scamp}. Absolute and relative
astrometric calibration is performed using SCAMP as well.  Resampling
for the application of the astrometric solution and final image
coaddition has been obtained using SWARP \citep{swarp}.

We note that the lack of a significant number of standard stars with 
$(g-i)<0.0$ mag makes it difficult to constrain the photometric 
calibration in the colour range -0.7$<(g-i)<$0.0 mag. For this reason,
the photometric calibration for stars with the bluest colours, 
corresponding to the young stellar populations in the SMC, could
be less accurate than for stars in the colour range
0.0$<(g-i)<$2.5 mag. We are working to solve this problem by observing
a significant  number of  blue stars in the Stripe 82 standard field \citep{Ivezic2007}.
These observations will allow us to accurately constrain the colour terms in the
calibration  equations. 
In the meantime, we can check quantitatively the goodness of our
photometry by comparing our data with those from the MCPS survey. To this aim, we have
first to transform the MCPS $BVI$ photometry into SDSS $g,i$. This can
be achieved by adopting the \citet{Jordi2006}
results\footnote{\bf For ease of use, we adopted \citet{Jordi2006}'s transformation
  equations in the form available at 
http://www.sdss3.org/dr8/algorithms/sdssUBVRITrans\\form.php.} for
population I objects, as the metallicity of the SMC is closer to
population I than population II stars. We note that these transformations are
nominally valid for $(V-I)<1.8$ mag, i.e.  $(g-i)<2.1$ mag, however, it
is not clear which is  the validity boundary towards blue colours. Secondly, we have matched the
STEP and MCPS catalogues accepting stars within 1$^{\prime\prime}$ and
with S/N$\geq$20 in each band (for both surveys). The result of this comparison
for tiles 3\_7 and 4\_6  is shown  in Fig.~\ref{comparison}. These
figures show that there is a fairly good agreement between the
two photometries. Even if a significant scatter ($\sim$0.08
mag) is present the formal average differences (in the sense
STEP-MCPS) are: $\Delta g$(3\_7)$\sim -0.004$ mag; $\Delta i$(3\_7)$\sim
-0.018$ mag; $\Delta (g-i)$(3\_7)$\sim +0.014$ mag;  $\Delta
g$(4\_6)$\sim +0.014$ mag; $\Delta i$(4\_6)$\sim
-0.014$ mag; $\Delta (g-i)$(4\_6)$\sim +0.027$ mag.
The maximum deviation is less than 3\% in color in tile 4\_6,
confirming the absence of significant photometric errors on the calibration.

   \begin{table}
  \begin{center}
     \caption[]{Observing Log for tiles 3\_7 and 4\_6.}
        \label{ObsLogstep}
        \begin{tabular}{c c c  c c}
           \hline 
           \hline 
          \noalign{\smallskip}
tile &    filter &       Date (UT)   & Exp. Time (sec) & seeing ($^{\prime\prime}$) \\
           \noalign{\smallskip}
           \hline 
           \noalign{\smallskip}
3\_7   &   $g$  &    2012/10/13            &   125s; 2600s   & 0.95 \\      
3\_7   &   $i$   &    2012/10/13            & 125s; 2600s   & 0.85 \\      
3\_7   &   $g,i$ Cal.   & 2012/08/14   & 45s & 2.1,1.8 \\      
4\_6    &  $g$   &         2012/09/20       & 125s; 2600s   & 1.05 \\      
4\_6    &  $i$    &     2012/08/9               &    125s; 2600s  & 0.85 \\      
4\_6    &  $g,i$ Cal.    &   2012/08/14       &     45s & 1.7,1.2 \\              
         \noalign{\smallskip}
           \hline 
	\end{tabular}
	\end{center}
  
  \end{table}

\begin{figure*}
\centering 
\includegraphics[width=10cm]{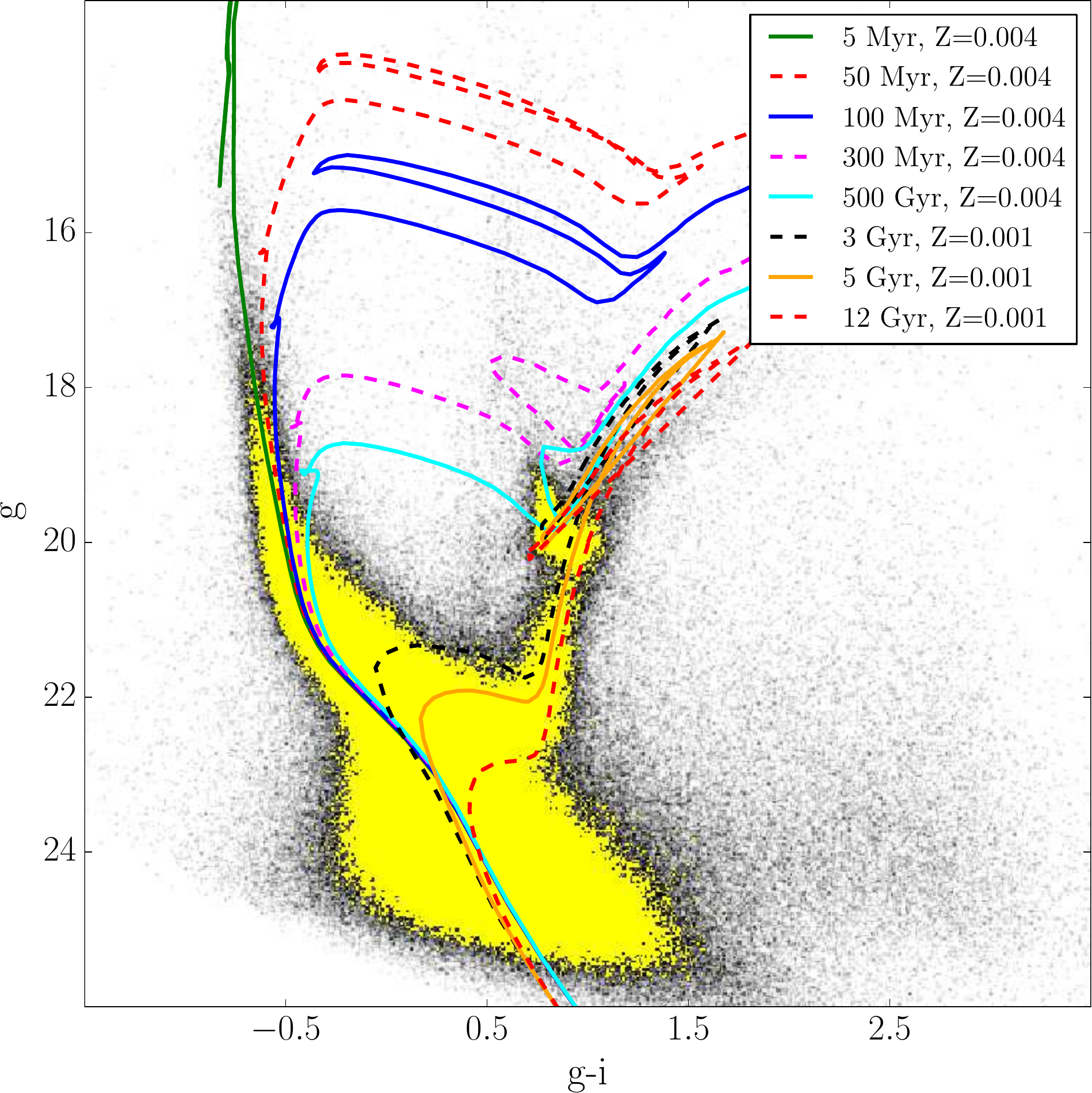}
\caption{CMD of tile 3\_7 with overlaid \citet{marigo08} stellar 
  isochrones for metal abundance Z=0.004, ages 5 Myr (green continuos 
  line), 50 Myr (red dashed line), 100 Myr (blue continuous line), 300 
  Myr (pink dashed line) and 500 Myr (cyan continuous line);   
  Z=0.001,  ages 3 Gyr (black dashed line), 5 Gyr (orange 
  continuous line), and 12 Gyr (dashed red line). Assumed distance modulus and 
  reddening E(B-V) are 18.9 mag and 0.08 mag, respectively.}
\label{f37_iso}
\end{figure*}

 \begin{table}
   \begin{center}
      \caption[]{Absolute photometric calibration for  tiles 3\_7 
        and 4\_6 (rows 1,2 and 3,4, respectively). The colour term is
        with respect to $(g-i)$. These coefficients are accurate for
        $(g-i)>0$ mag. The formal uncertainty on zero-points (ZP) and colour
        terms are also shown.
The date format is dd mm yyyy. }
         \label{zp}
         \begin{tabular}{cc@{}c  c@{}c@{} c}
            \hline 
            \hline 
          \noalign{\smallskip}
   tile & band & ZP & Col. term  & extinction & date \\
           \noalign{\smallskip}
     &  & mag & mag & mag/airmass\\
           \noalign{\smallskip}
           \hline 
           \noalign{\smallskip}
3\_7 &  g & 24.767$\pm$0.016      & 0.016$\pm$0.013 &    0.18 & 13 09 2012\\
3\_7  &    i &  24.094$\pm$0.014  & -0.003$\pm$0.011 & 0.043   & 13 09 2012\\
4\_6  &    g &  24.832$\pm$0.015  & 0.015$\pm$0.012  & 0.18   & 20 09 2012 \\
4\_6  &    i &  24.107$\pm$0.007  & -0.003$\pm$0.006 & 0.043   & 09 08 2012\\
            \noalign{\smallskip}
            \hline\\
	\end{tabular}
	\end{center}
   \end{table}

\subsection{Photometry}

Our fields present different levels of crowding. The SMC body, in
particular, is heavily congested with stars. This requires the use of
Point Spread Function (PSF) photometry. We employed the
package DAOPHOT IV/ALLSTAR \citep[][]{stetson87,stetson92} and the PSF
function was left to vary across the FoV. Usually a first or second
order variation of the PSF with radial position was sufficient to
recover star shapes even at the edges of the tiles.

At the end of the pre-reduction procedure we were left with three
images for each filter and deep tile.  Two are mosaics resulting from
the stack of the short and long exposures, respectively (see
Tab.~\ref{observingStrategy}), the third image is the single shot
acquired to produce secondary standard stars. Each image was measured 
separately. For each filter the output short and long exposure files
were matched to adjust the residual photometric zero point difference
(usually of the order of 0.01-0.015 mag) by averaging the photometry of
the stars in common. This match was performed in $\alpha$ and
$\delta$, which were calculated from the ALLSTAR x,y physical
coordinates using the World Coordinate System (WCS) of the images and the package {\it
  xy2sky}\footnote{http://tdc-www.harvard.edu/wcstools/xy2sky/}.  For
the match we used the {\it
  STILTS}\footnote{http://www.star.bris.ac.uk/~mbt/stilts/} package
setting a tolerance in both $\alpha$ and $\delta$ of
0.25$^{\prime\prime}$, i.e. slightly more than 1 pixel. The resulting
$g,i$ catalogues were matched in the same way, allowing us to obtain a
unique catalogue including all the photometry for each tile.

\begin{figure*}
\centering 
\includegraphics[width=10cm]{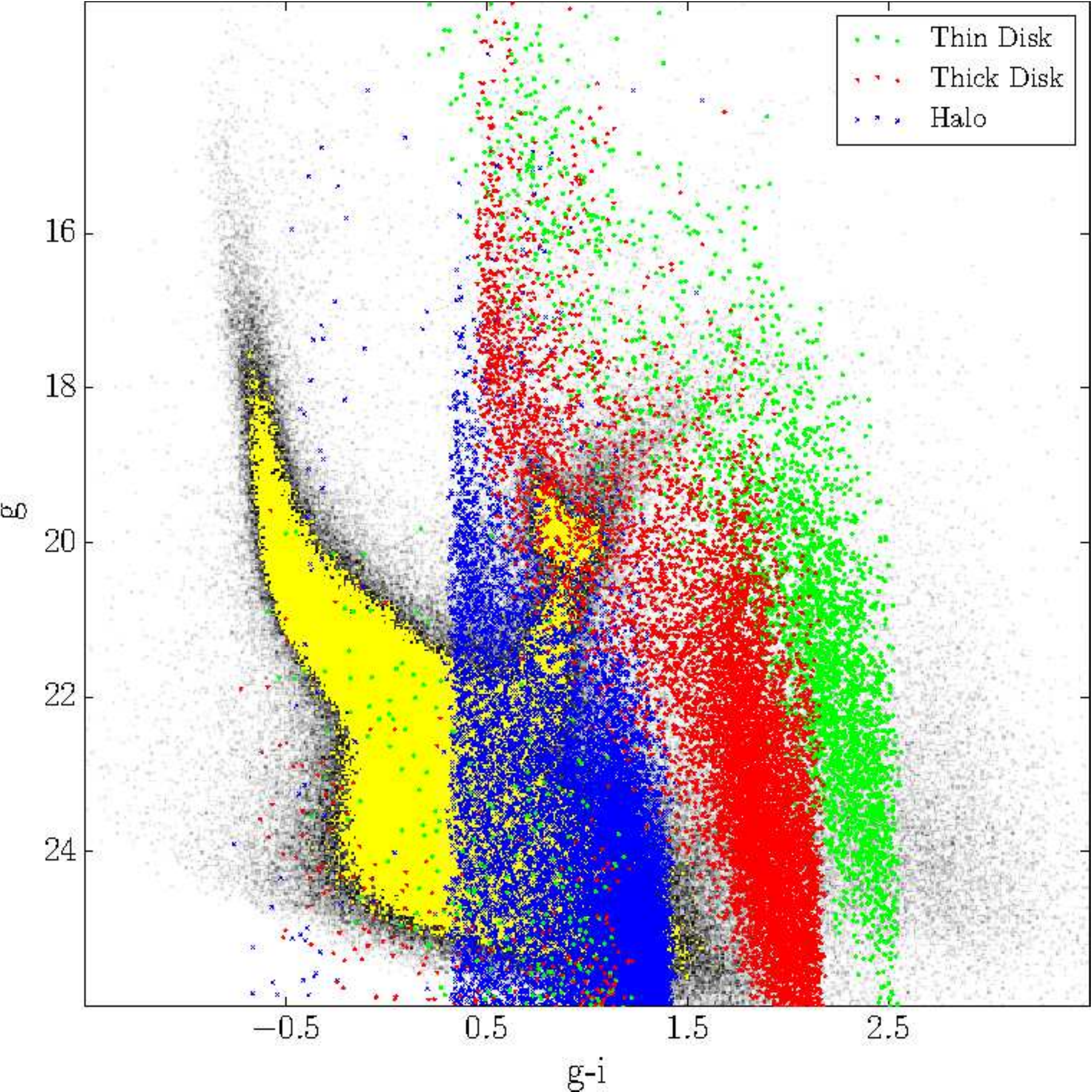}
\caption{CMD of tile 3\_7 with overlaid a simulation of MW 
  contamination expected in the FoV. Halo stars are in blue, Thick 
  disk stars in red and thin disk stars in green.}
\label{f37_simul} 
\end{figure*}

Similarly, the pair of $g,i$ single exposures acquired with the
purpose of providing secondary standards, were matched. For all the
stars in common we then applied the colour terms listed in Tab.~\ref{zp} 
(we recall that the
zero point is already photometrically calibrated by the VST--Tube
pipeline), obtaining the desired secondary standard stars.  

Finally, the last step of the whole procedure was the refinement of the
photometric calibration. This is achieved by matching the catalogue of
the secondary standard stars with that including all the photometry of
the tile and correcting for the possible difference in zero
points. Since most of the data have been acquired in ``clear''
conditions, corrections were found to be small, of the order
of a few percent.  

To show the quality of the photometry, we report in Fig.~\ref{rms} the
errors provided by ALLSTAR in $g$ and $i$ as a function of the magnitude for the nine
sub-frames in which we virtually divided  each tile with the aim of
identifying possible differences in the behaviour of the photometric
errors.



\subsection{Completeness}

A detailed estimate of the completeness of our photometry is a
fundamental step for an accurate reconstruction of the SFH 
\citep[see e.g.,][and references therein]{tosi2001}.
We followed the usual procedure of adding artificial stars to
the images and calculating the percentage of the recovered stars. More
in detail, the steps of our procedure were the following:

\begin{itemize}

\item
The colour, the magnitude  and the relative frequency of the stars to be inserted in the images
were calculated on the basis of a synthetic CMD roughly representing
a population characteristic of the SMC and Bridge.  
The range of magnitude taken into account was 14$<$$g$$<$26.5 mag
and the total number of artificial stars  was about 2 millions for
each tile, distributed along many different completeness experiments
(see next point). 

\item
To avoid self-crowding, we sub-divided the images in virtual  boxes 
70x70 pixels in size each. One artificial star is placed inside
each box at random position, so that we can place several thousands of
artificial stars avoiding the problem of self-crowding.

\item
A star is considered as recovered if it is measured both in $g$ and
$i$ and if its position and magnitude are returned within
0.25$^{\prime\prime}$ and 0.75 mag of its input values, respectively.

\end{itemize}

The result of this procedure is shown in Fig.~\ref{completeness} for
the two tiles presented in this paper. To verify the presence of
possible differences in completeness within each image, the tiles were
divided into nine sub-frames 20$^{\prime}$$\times$$21^{\prime}$
each. The completeness in $g$ and $i$ are represented with different
colours.  The figure shows that the magnitude level corresponding to
50\% completeness is brighter ($g\sim23.5$ mag) for tile 4\_6, that is
much more crowded than tile 3\_7 ($g\sim24-24.5$ mag). It is
instructive to compare these values with those obtained with HST by
\citet{sabbi09} for the fields SFH9 and SFH4, included in the tiles
3\_7 and 4\_6, respectively. Indeed, an inspection of their Figure 3
reveals that for HST data 50\% completeness was reached at
$V$$\approx$26.2 mag, $I$$\approx$26 mag and $V$$\approx$24 mag,
$I$$\approx$25.2 mag for fields SFH9 and SFH4, respectively. Even if
the $g$ and $V$ magnitudes cannot be directly compared, we can roughly
conclude that VST data become severely incomplete at about 1-1.5 mag
brighter than with the HST photometry.
   
Artificial stars also provide a realistic estimate of the
  photometric error, hence the best age resolution achievable within
  STEP. In the range $22.5<g<23$ mag (approximately the magnitude of the
  oldest TOs in the SMC), tile 4\_6 is characterised by a photometric
  accuracy in g magnitude of about 0.2 mag. This translates into a
  maximum precision of $\sim$ 17\% in the derived ages
  \citep[see][]{vandenberg85}, which is equivalent to an age error of 1.7
  Gyr at 10 Gyr. For tile 3\_7, however, crowding conditions are less
  severe and the age error at 10 Gyr drops at about 1 Gyr. Although
  these performances are not comparable with those reached with HST,
  they are superior to the performances of the MCPS, where the maximum
  age resolution at 10 Gyr is worse than 3 Gyr.

\section{The stellar populations in the central and Wing regions}

As planned by our observing strategy, the first STEP fields for which 
data acquisition was completed were tile 4\_6, around the very young 
cluster NGC\,346, and tile 3\_7 in the wing. As described in Section 
2.3, field 3\_7 includes our HST/ACS field SFH9, while field 4\_6 
covers our HST/ACS pointing on NGC\,346 and also overlaps with the ESO 
Public Survey VMC tile SMC 5\_4.

\subsection{Tile  3\_7}

\subsubsection{CMD}

The CMD of all the stars measured in tile 3\_7 is displayed in
Fig.~\ref{f37_iso}.  There are a number of stellar evolution phases
easily identifiable in the CMD, including a well defined and extended
MS, a prominent RC (visible at $0.75<g-i<1$ mag and $19.7<g<20.2$ mag)
and a blue loop (BL; mostly visible at $0.7<g-i<0.9$ mag and
$19<g<19.5$ mag).

\begin{figure*}
\centering 
\includegraphics[width=15cm]{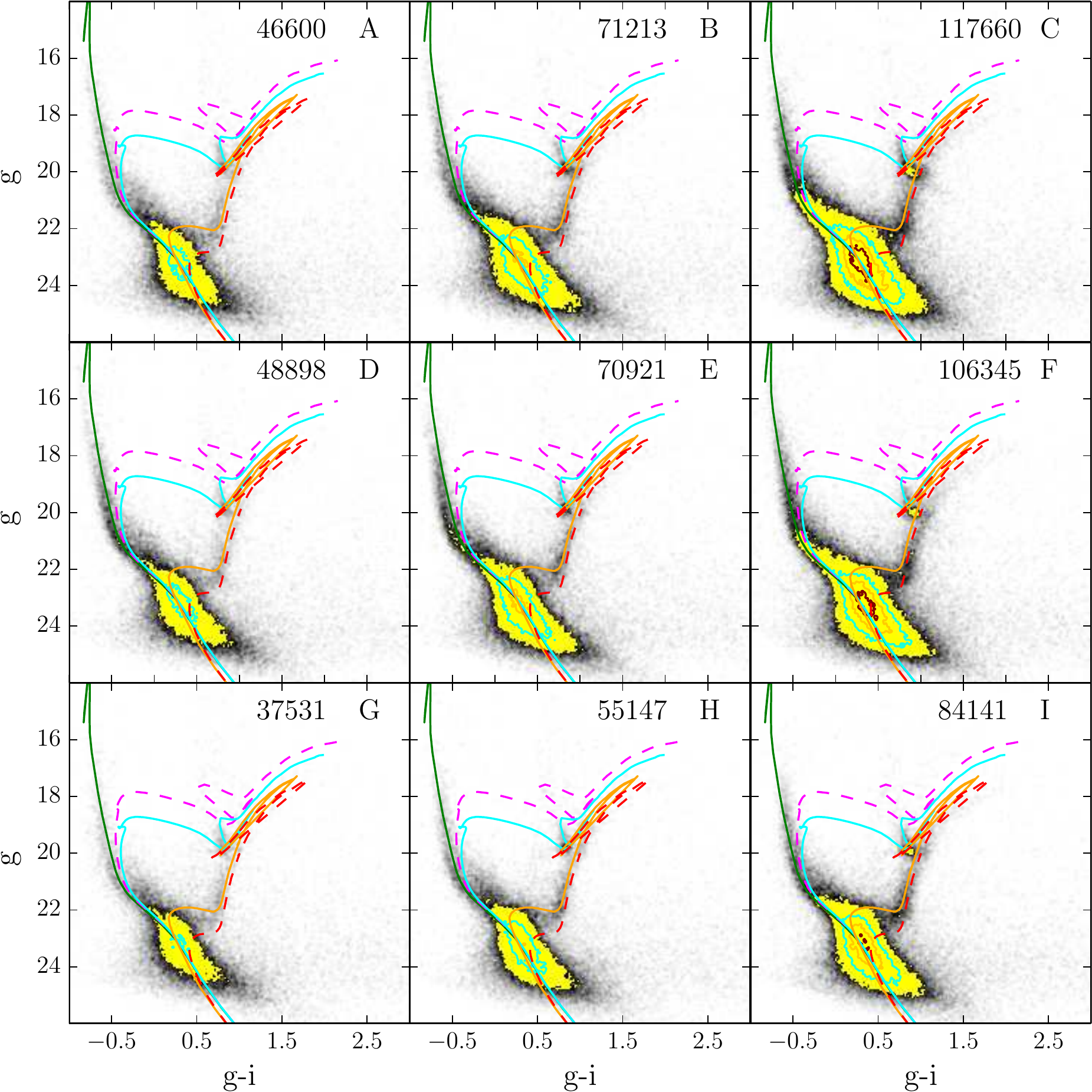}
\caption{ CMDs for nine $20^{\prime}\times21^{\prime}$ arcmin  
  subregions of tile 3\_7.  The isochrone ages and metallicities are  
  the same as in Figure \ref{f37_iso}. Assumed distance modulus and  
  reddening E(B-V) are 18.9 mag and 0.08 mag, respectively.}
\label{9cmd_f37} 
\end{figure*}
 
There are also contaminants. Figure \ref{f37_simul} shows the data CMD
with overlaid a simulation\footnote{Photometric errors and
  incompleteness are not included.} of the MW foreground expected in
tile 3\_7, according to the Galactic model for star counts described
in \citet{castellani02} and \citet{cignoni07}. Different colours represent the
three major galactic components, namely Halo (blue), Thick Disk (red)
and Thin Disk (green). According to  this comparison we suggest that
in the CMD of Fig. \ref{f37_iso}, the group of stars
in the colour range $1.75<g-i<3$ mag and $g>20$ mag and the vertical
sequence at $g-i\approx 0.6$ mag and $g<18$ mag are likely foreground
MW stars along the line of sight. However, it is also noteworthy that
simulations do not show MW stars redder than $g-i \approx 2.5$ mag,
while observations reach $g-i \approx 3$ mag and beyond. Although
there are known theoretical problems to reproduce observed stellar
colours at such low temperatures, we suggest that part of the mismatch
is due to background galaxies. Indeed, using near-infrared VISTA
observations for LMC fields, \citet{rubele12} find unequivocally that
background galaxies are important contaminants, dominating star counts
for $Y-K_\mathrm{s}>$2 mag and $K_\mathrm{s}>$17 mag, while MW
foreground stars are concentrated at $Y-K_\mathrm{s} \approx$ 1-1.5 mag.

In Fig. \ref{f37_iso} isochrones of different ages and metallicities
from \citet{marigo08} are overlaid to the CMD. The metallicity of the
youngest isochrones is assumed to be Z=0.004, which is consistent with
spectroscopic derivations from HII regions in the SMC \citep[see
  e.g.,][]{Russel1989,Kurt1998} and from stellar abundances of very
young stars \citep[see
  e.g.,][]{Gonzalez1998,Hunter2007,Hunter2009,Bouret2013}, while the
metallicity of the older isochrones (Z=0.001) is chosen to best fit
the RGB \citep[see e.g.,][]{Butler1982,Haschke2012b}. The assumed
distance modulus is $(m-M)_0$ = 18.90 mag \citep[see e.g.,][]{harries03},
while the reddening value is E(B-V)=0.08 mag. 

\begin{figure*}
\centering 
\includegraphics[width=12cm]{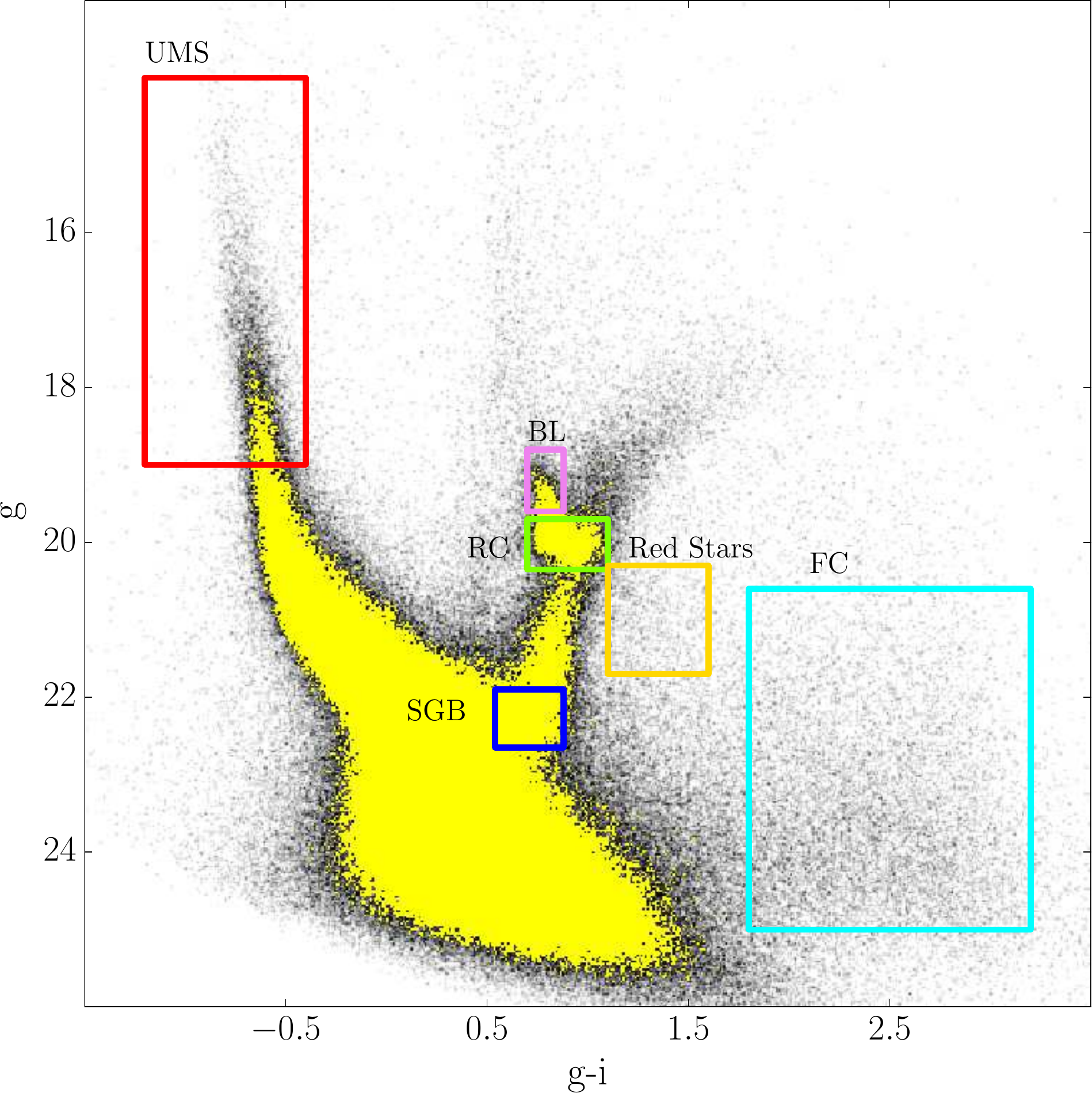}
\caption{Tile 3\_7. CMD regions used to isolate UMS, BL, RC, SGB,
  reddened RC stars and FC sources.}
\label{cmd_sel_f37} 
\end{figure*}

The presence of bright MS stars, BL and RC stars provides
circumstantial evidence that tile 3\_7 has been actively forming stars
for the last few Gyrs. Moreover, the conspicuous number of upper-MS
(UMS) stars between the 5 Myr and 100 Myr isochrones suggests that the
region went through a recent intense star formation phase producing
such massive stars. Looking back in time, the next relevant feature is
a rather short BL, roughly consistent with a 400-500 Myr old
population, while the paucity of BL stars brighter than $g=19$ mag 
suggests a reduced star formation 100-300 Myr ago. The RC is roughly
round and well populated, which implies a continuous activity at
intermediate epochs. On the other hand, there is no evidence for an
extended or even red HB, suggesting a very low star formation rate
earlier than 10 Gyr ago. Overall, these results compare well with
previous investigations of the region.  Using deep HST/ACS data,
\citet{cignoni13} studied the detailed star formation of a field (SFH9)
 included in tile 3\_7. Their results show a broadly constant
intermediate age activity (1-7 Gyr ago), with a recent burst of star
formation overimposed, and very modest activity in the first few Gyr.

\begin{figure*}
\centering 
\includegraphics[width=14cm]{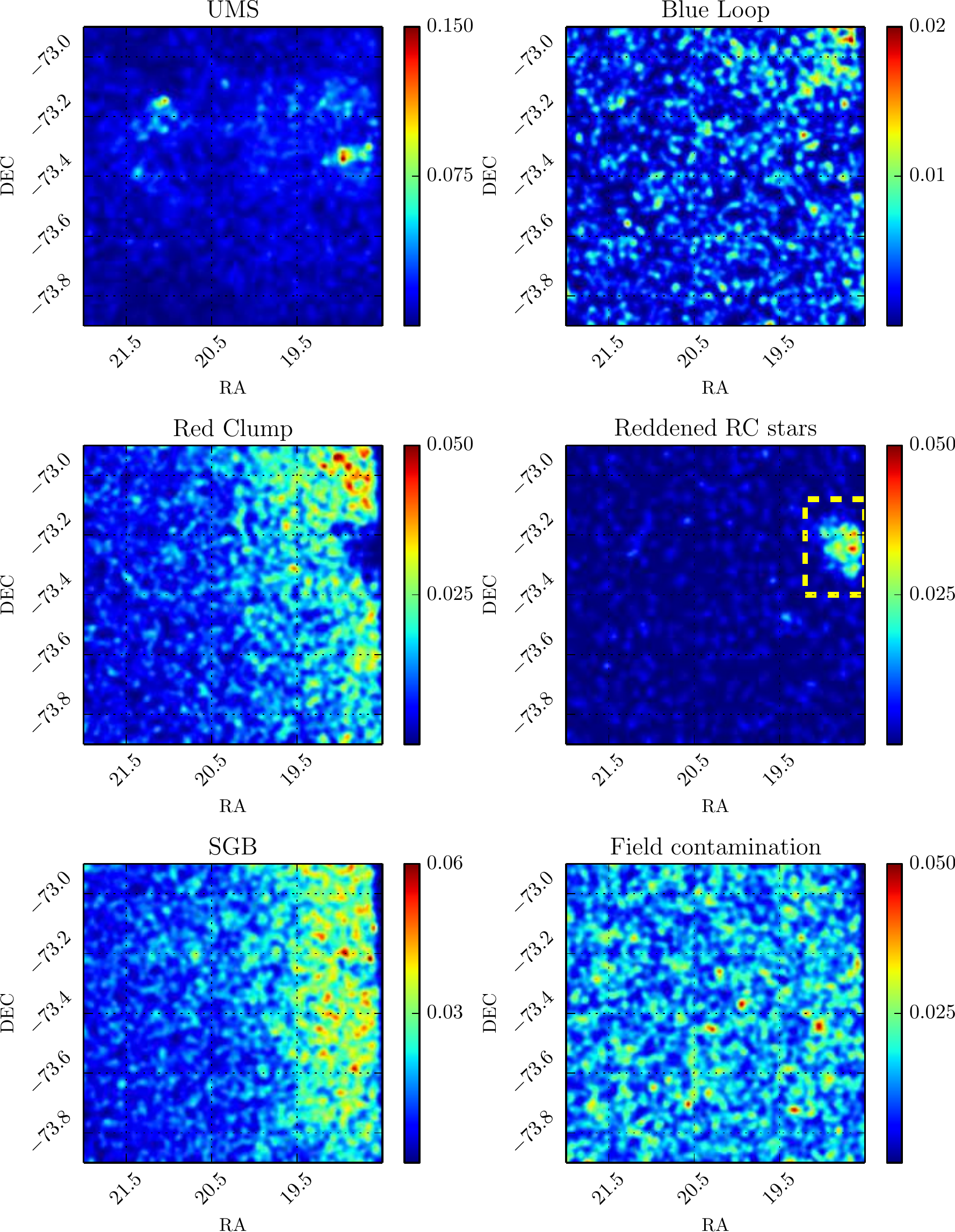}
\caption{ Tile 3\_7. Spatial distributions of labeled CMD selections 
  (see Fig. \ref{cmd_sel_f37}). Maps are produced by calculating a 2D 
  histogram of stellar positions and then smoothing the result with a 
  gaussian kernel. The yellow dashed box (in the middle-right panel) 
  highlights a subregion where reddening is above the average (see 
  Fig. \ref{cmd_red}).}
\label{maps_f37} 
\end{figure*}

In order to better visualize the variation of extinctions and crowding
conditions, we divided tile 3\_7 into nine sub-regions
$20^{\prime}\times21^{\prime}$ arcmin large, and compared their CMDs
(see Fig. \ref{9cmd_f37}) to the same set of isochrones. The
old isochrones bracket well all the observational RGBs, except in
subregions C and F where the foreground reddening is probably higher.

On the other hand, the blue edge of the MS appears often bluer than
our youngest models (see e.g. subregion E). Adopting a slightly lower
metallicity or reddening could mitigate this effect, but would cause
other inconsistencies. In fact, the adopted metallicity (Z=0.004) provides
excellent fits to the MS derived from HST data \citep[see
  e.g.,][]{cignoni12,cignoni13}. Moreover, younger populations are
generally more reddened than older ones \citep[see
  e.g.,][]{Zaritsky2002}, while the data would require the
opposite. We do not elaborate further on this issue because of
possible problems in the calibration of blue stars.  As discussed in
Sect.~\ref{prereduction} this aspect will be improved in future
papers. Here our main aim is only to show the full potential of the
STEP survey.


\subsubsection{Spatial distribution}

The spatial distribution of stars in different evolutionary stages
yields important information on the star formation processes over the
region. We have counted stars in different age regions of the
CMD (see Fig.~\ref{cmd_sel_f37}), sampling the UMS (red box), RC
(green box), BL (pink box) and sub-giant branch (SGB, blue box)
stars. Field contamination (FC; interlopers from the MW and background
galaxies) and highly reddened RC stars are sampled with the cyan and
yellow boxes, respectively. Fig. \ref{maps_f37} shows the spatial
distributions of these different groups of stars.

From the spatial point of view, we note that populations grow more
compact as one moves toward younger ages. In particular:

\begin{itemize}

\item UMS stars (top left panel in Fig. \ref{maps_f37}) appear
  clumpy and irregularly distributed. Two major concentrations are
  visible at ($\alpha=21.1$, $\delta=-73.2$) and ($\alpha=18.9$,
  $\delta=-73.3$). In terms of age, this implies that most of the
  ongoing and recent (i.e. in the last 100 Myr) star formation is
  concentrated in these locations, while the activity is stagnant in
  the rest of the field;

\item BL stars (top right panel in Fig. \ref{maps_f37}) appear more
  evenly spatially distributed than UMS stars, but still a bit
  clumpy. However, given the paucity of these objects compared to
    UMS stars (ten times more numerous), it is difficult to assess the
    significance of the various structures, except the mild gradient
    of increasing concentration toward the upper-right corner of the
    FoV, which is clearly overdense with respect to the opposite
    corner. Physically, this implies that either star formation
  younger than 500 Myr and older than few Myr (the minimum age of BL
  stars) is occurring on extended scales, not just in few clumps like
  the UMS, or that stars have been able to disperse during this length
  of time.

\item Overall, RC stars (middle-left panel) follow the spatial
  distribution of BL stars, but are more evenly distributed along the
  $\delta$ axis. This reflects the fact that RC stars are intermediate
  age stars (1-8 Gyr ago), so they have had sufficient time to diffuse
  throughout the SMC. The lack of RC stars (``hole'') around
  ($\alpha=18.8$, $\delta=-73.25$) is likely due to an extended
  cloud of obscuring matter \citep[actually the ``hole" is part of a
    large region showing large extinction - see Fig. 4
    in][]{Haschke2011}. Indeed, when we look at the map (middle-right
  panel) of stars taken from the yellow box in the CMD of
  Fig. \ref{cmd_sel_f37} we find a clear overdensity at the location
  of the hole, hence suggesting that the missing RC stars are actually
  normal RC stars moved to the yellow box because of higher
  reddening. As a further support of this interpretation,
  Fig. \ref{cmd_red} shows the CMD of all stars near the hole (yellow-dashed box 
   in the middle-right panel of
  Fig. \ref{maps_f37}). Its elongated RC, broad RGB and upper MS are
  unique signatures of a large differential reddening. However, the
  total number of highly reddened stars is less than 1\% of the stars in
  the entire tile, hence they are unnoticeable in the CMD of
  Fig. \ref{f37_iso};

\begin{figure*}
\centering 
\includegraphics[width=12cm]{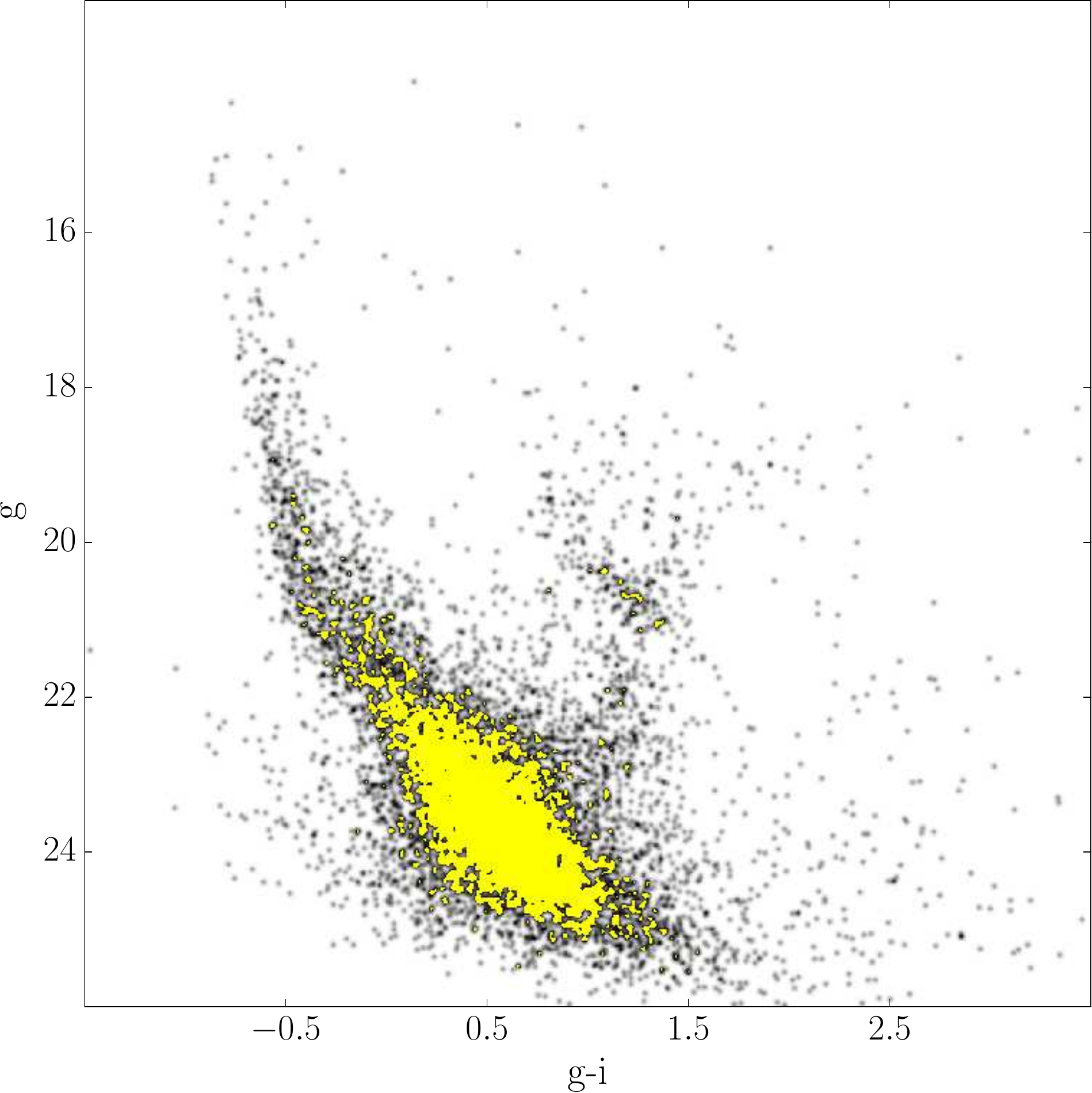}
\caption{CMD for reddened stars in the tile 3\_7  (yellow-dashed box in the 
  middle-right panel of Figure \ref{maps_f37}.)  }
\label{cmd_red} 
\end{figure*}

\item The distribution of SGB stars (bottom-left panel) shows a net
  gradient of increasing concentration from left to right,
  without significant excess at the upper right corner (where BL
  stars are more concentrated). This mainly reflects the increasing
  number of star counts moving towards the SMC centre;

\item Field contaminants (bottom right panel) are uniformly distributed over
  the FoV. The only few visible overdensities are likely due to MS
  stars scattered into the FC box because of the reddening.

\end{itemize}

\subsection{Tile 4\_6}

\subsubsection{CMD}

\begin{figure*}
\centering 
\includegraphics[width=13cm]{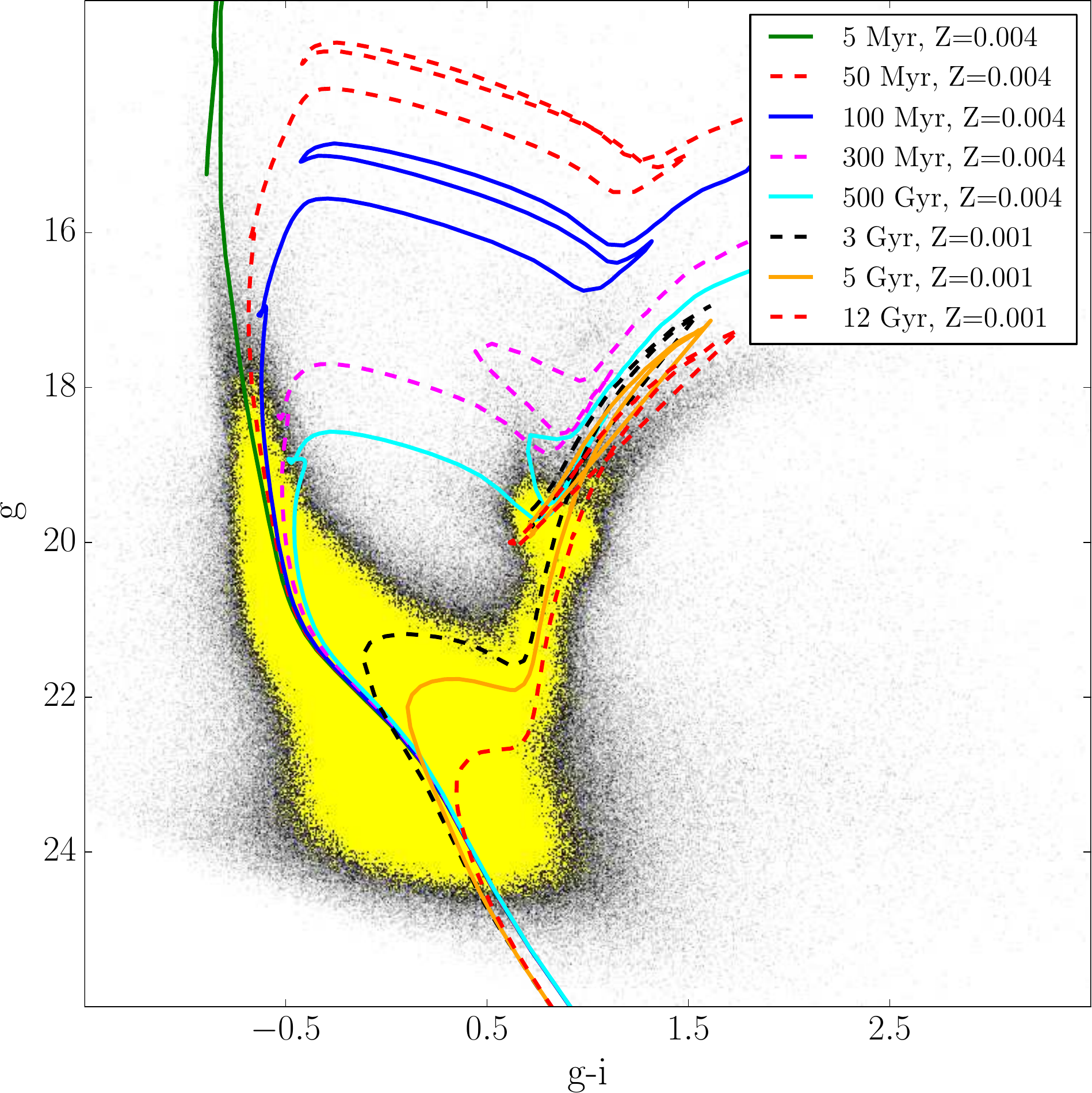}
\caption{CMD of tile 4\_6. The isochrone ages and metallicities are
  the same as in Figure \ref{f37_iso}.  Assumed distance modulus and
  reddening E(B-V) are 18.9 mag and 0.04 mag, respectively.}
\label{cmd_iso_f46} 
\end{figure*}

Tile 4\_6 hosts over four times more stars than tile 3\_7. As a
consequence crowding is more severe and the corresponding CMD (Figure
\ref{cmd_iso_f46}) much shallower (but still one mag deeper than the
oldest MSTO) and broader. In terms of stellar populations, there are
two striking differences from 3\_7's CMD (see Figure \ref{f37_iso}):
1) 4\_6's CMD shows a prominent RGB bump, just above the RC, while
this feature is missing in tile 3\_7; 2) the morphology of the RC is rather
elliptical in the 4\_6 CMD, while it shows a clear protrusion towards
brighter magnitudes in tile 3\_7. In terms of age/metallicity the evidence
of an RGB Bump brighter than the RC is a clear indication that
intermediate and old star formation took place at relatively low
metallicity (Z=0.001 or less) in tile 4\_6. However, its apparent lack
in the tile 3\_7 CMD could be just due to the lower number of stars. On
the other hand, the populous RC protrusion in the tile 3\_7's CMD represents a
significant difference from tile 4\_6, since the latter is globally much
more populated. Considering that the RC protrusion is likely populated
by objects at the transition between RC and BL phases, therefore by stars
with ages between 500 Myr and 1 Gyr, our conclusion is that tile 3\_7
has been relatively more active than tile 4\_6 at these epochs. If we
take into account that tile 3\_7 is in the Wing region, at the edge
 of the Bridge connecting the SMC to the LMC, it is
tantalising to relate the enhanced star formation of tile 3\_7 with an
SMC/LMC interaction.

\begin{figure*}
\centering 
\includegraphics[width=15cm]{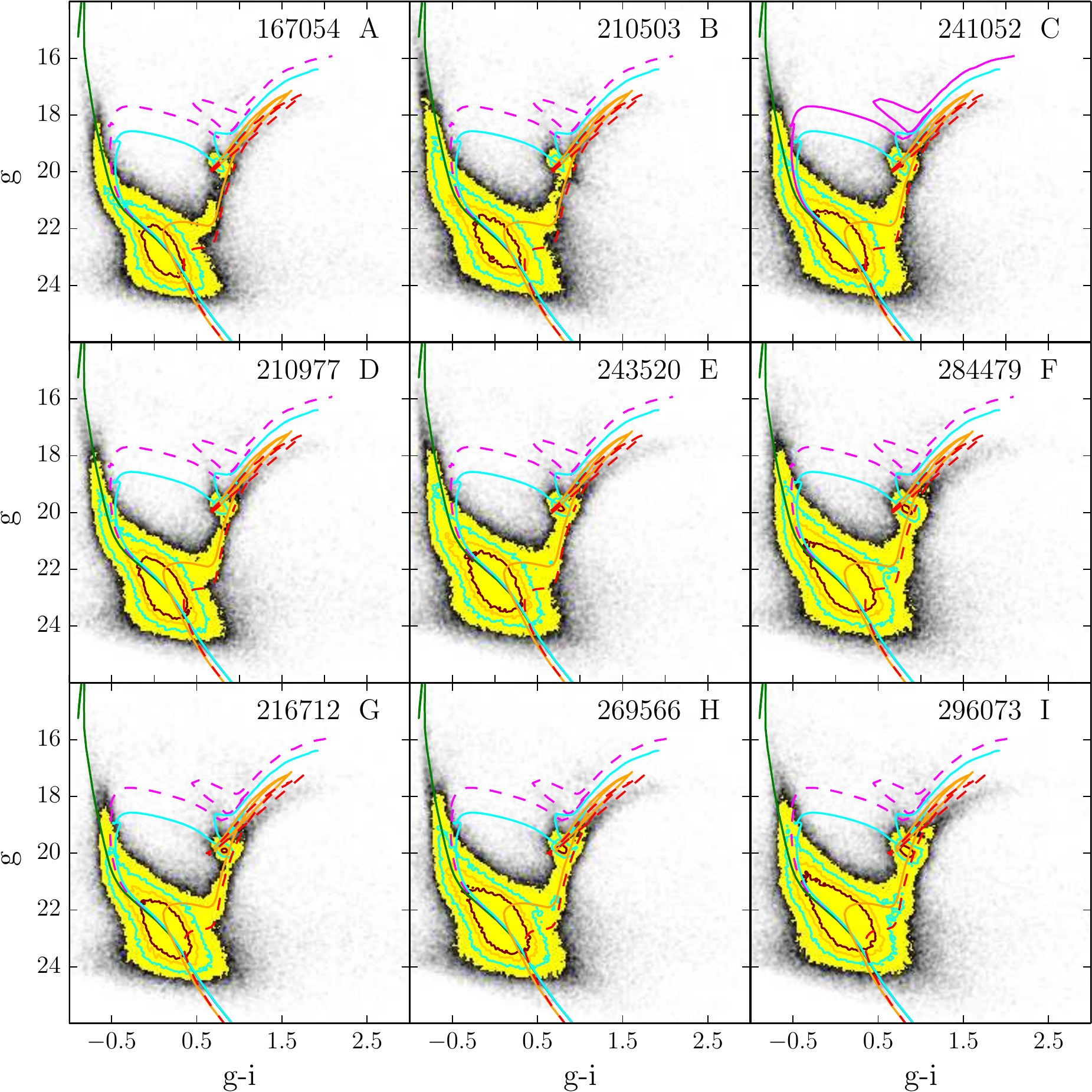}
\caption{ CMDs for nine $20^{\prime}\times21^{\prime}$ arcmin
  subregions of tile 4\_6.  The isochrone ages and metallicities are
  the same as in Figure \ref{f37_iso}.  Assumed distance modulus and
  reddening E(B-V) are 18.9 and 0.04, respectively.}
\label{9cmd_f46} 
\end{figure*}

As for tile 3\_7, the comparison with stellar isochrones shows an overall
good agreement. We derive a reddening E(B-V)=0.04, which is lower than
in tile 3\_7. As in tile 3\_7,  the chosen isochrones do
not bracket the whole colour distribution spread which characterises both the MS and
RGB. In particular, the observational CMD extends much further to the
red than the models, hence suggesting that a minor fraction of stars in tile 4\_6
are either more reddened or more metal rich than the average of the tile.
Once again, the division in sub-regions helps to disentangle the
problem (see Fig.~\ref{9cmd_f46}). Indeed, the sub-regions where
the RGB is redder than models (for example, subregion F and I) are 
those whose MS is globally redder, which is more consistent with a
reddening effect than a metallicity difference.

Concerning the MS, it is worth noticing that different subregions can
show very different morphologies. For example, the CMD of subregion G
shows a clear turn-off consistent with a star formation burst 100 or
200 Myr ago, while subregions like B show a much more extended MS,
probably consistent with an activity extending to a few Myr ago.
Interestingly, most of the regions show signs of the two events,
suggesting that the star formation burst in one subregion could have
triggered a burst in a neighbour subregion, as in the scenario derived
by \citet{Vandyk1998} and \citet{dp02} for Sextans A.

\begin{figure*}
\centering 
\includegraphics[width=14cm]{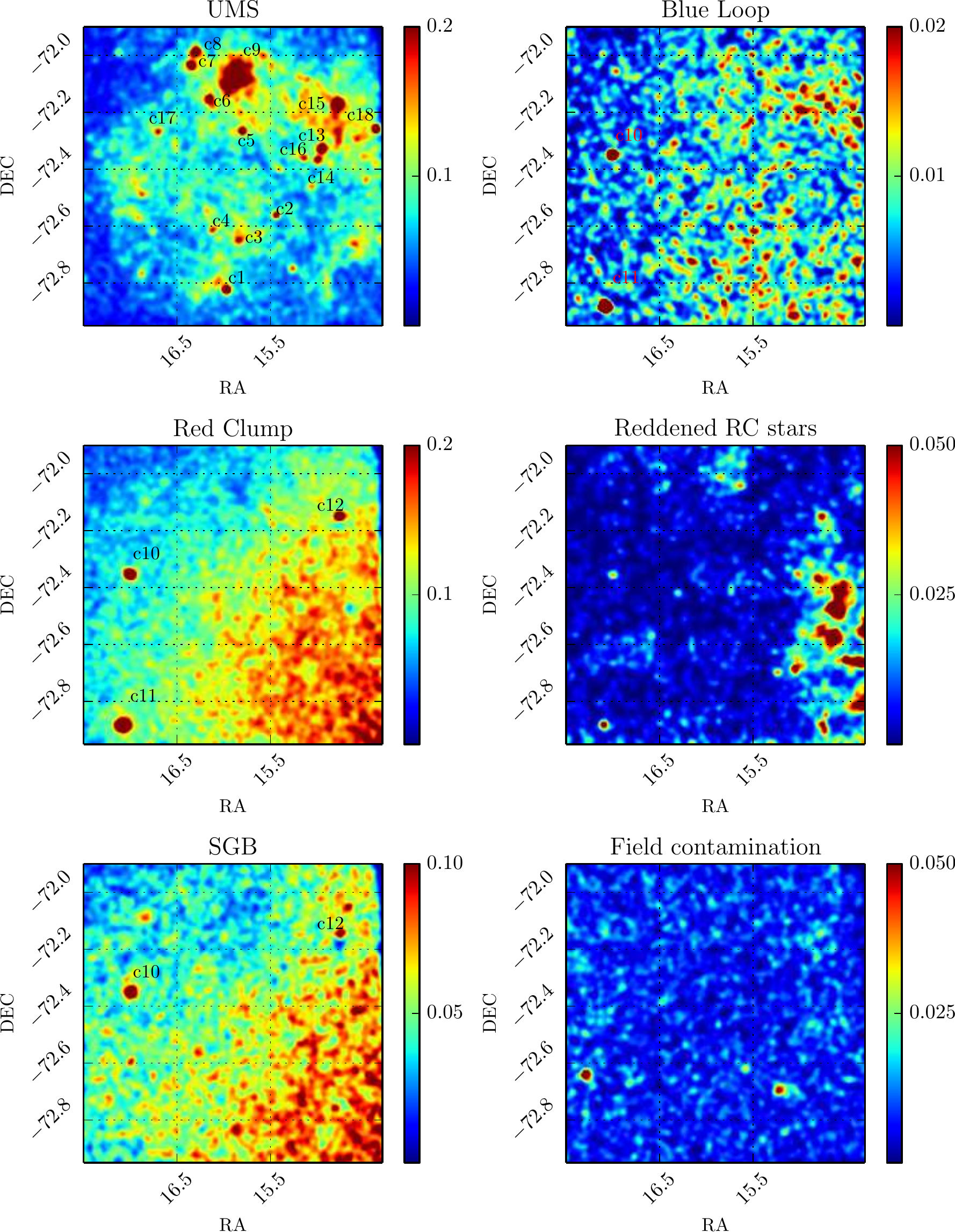}
\caption{Tile 4\_6. Spatial distributions of labeled CMD 
  selections. Labels c1 to c18 indicate clusters identified by eye.}
\label{maps_f46} 
\end{figure*}

\begin{table*}
\caption{Cross-Identification of the over-densities found in
  Fig.~\ref{maps_f46}. From left, the different columns show: the
  cluster ID of  Fig.~\ref{maps_f46}; the evolutionary phase in which
  they where identified; the equatorial coordinates; the SIMBAD name
  (L=\citet{Lindsay1958}; H86=\citet{Hodge1986};BS=\citet{Bica1995};B=\citet{Brueck1976});
  the classification (NA=Nebula-Association; C=Cluster;
  A=Association), the major and minor axis according to
  \citet{bic08}; the age and the reference for the age estimate: 
a) \citet{chiosi06}; b) \citet{Glatt2010}; c) \citet{Rochau2007}; d) \citet{glatt08b}. 
}
\label{crossIdClusters}
\begin{center}
\begin{tabular}{cccccccccc}
\hline
\hline
\noalign{\smallskip} 
ID & Ev. Phase & RA (J2000) & DEC (J2000) & Name & Class. & Major Axis& Minor Axis & log(Age) & Age Ref. \\
    &                  & deg  & deg  &  &  & arcmin & arcmin & Gyr &  \\
\noalign{\smallskip}
\hline
\noalign{\smallskip} 
c1      &  UMS    & 15.9708  &  -72.8261  &   NGC\,376       &          C    & 1.8        &  1.8   &7.50          &    b       \\
c2      &  UMS    & 15.4375  &  -72.5644  &   L66          &              C    & 1.1         & 1.1   &7.40      &        b   \\
c3      &  UMS    & 15.8458  &  -72.6517  &   B115         &             C    & 0.9         & 0.8   &7.60       &        b   \\
c4     &  UMS    & 16.0583  &  -72.6469  &  BS118         &            A    & 1.3        &  0.9   &6.9         &      a     \\
c5      &  UMS    & 15.8     &  -72.2725  &     L70      &                  C   &0.75     & 0.75     & 7.60    &         b  \\
c6      &  UMS    & 16.15    &  -72.1606  &     L74        &                C    & 1.0     &   1.0   & 7.00     &          b \\
c7      &  UMS    & 16.3417  &  -72.0431  &   IC\,1624       &           C    & 0.9    &  0.9     & 8.35          &     b      \\
c8      &  UMS    & 16.2792  &  -71.9936  &   NGC\,395       &        NA  &  1.1   &  1.1     &   7.2             &   a        \\
c9      &  UMS    & 15.8708    &  -72.0567  &   NGC\,371        &     NA    & 4.2       & 3.8  &    6.7   &  a        \\
c10     &  BL,RC,SGB    & 16.9958  &   -72.3556 &  NGC\,416        &         C     &1.7        &  1.7   &9.80           &    d       \\
c11      & BL,RC    & 17.0792  &   -72.8842 &  NGC\,419        &         C     &2.8     & 2.8   &9.00               &    d       \\
c12      &  RC,SGB    & 14.775   &    -72.1508&   BS90         &              C    & 1.0        &  1.0  &9.6        &       c    \\ 
c13     &  UMS    & 14.95    &  -72.3339  &     IC\,1611     &              C    & 1.5       &   1.5   &8.20      &       b    \\
c14      &  UMS    & 14.9875  &  -72.3733  &  H86-186       &         C    & 0.6         & 0.6   &8.20           &      b     \\
c15     &  UMS    & 14.7708  &  -72.1769  &   NGC\,346       &         NA   & 8.5      &  8.5   &7.2              &  a         \\
c16      &  UMS    & 15.1417  &  -72.3656  &  L63           &              C    & 0.85        & 0.85  &7.80      &         b  \\
c17    &  UMS    & 16.7     &      -72.2736 &  L79        &                 C    & 0.9         & 0.9   &7.55   &        b   \\
c18     &  UMS    & 14.3792  &  -72.2644  &  L56           &             C     &0.95        & 0.95  &7.80       &        b   \\
\noalign{\smallskip}
\hline 
\end{tabular}
\end{center}
\end{table*}

\subsubsection{Spatial distribution}

Figure~\ref{maps_f46} shows how stars of different evolutionary phases
are distributed over tile 4\_6. Unlike tile 3\_7, it is clear that
4\_6 harbours many clusters and associations, probably because of the
generally higher activity of the region. As in tile 3\_7, the stellar
density of older populations (RC and SGB, middle-left and bottom-left
panels, respectively) increases smoothly towards the SMC centre (in
this case, the lower right corner of tile 4\_6), while the density of
younger populations (UMS and BL, top-left and top-right panels,
respectively) is very irregular and dominated by
inhomogeneities. Indeed, most UMS stars are found aggregated in
clusters/associations (identified clusters are indicated with labels
c1 to c18 in the maps in Fig.~\ref{maps_f46}; Table \ref{crossIdClusters} shows their
literature names and properties), with the most prominent ones
corresponding to NGC\,346 (c15 in the map) and NGC\,371 (c9).

We point out that the distribution of RC stars shows clusters as well
(c10, c11, c12), but none of them turns out to have a counterpart in
the UMS map. A simple explanation is that all agglomerates found in
the latter are younger than 100 Myr, hence too young to host RC
stars\footnote{But old enough to host BL stars. Indeed, clusters c10
  and c11 are clearly visible in the BL map.}. Vice-versa, clusters
detected only in the RC maps are necessarily too old to have UMS stars
still alive. It is also worth of notice that among clusters visible
using RC stars, namely c10, c11 and c12, only c10 and c12 are seen
also in the SGB map. Indeed, the c11 cluster is the well known
NGC\,419 (see Table \ref{crossIdClusters}), which is about 1 Gyr old,
hence the transition between the MS and the RGB phase, i.e. 
the SGB phase, is poorly populated (Hertzsprung Gap).  Note the excellent correspondance
between the literature ages reported in Table~\ref{crossIdClusters}
and the evolutionary phase adopted to detect the different c1-c18
structures.

Finally, the map of reddened RC stars (middle-right panel) resembles
that in tile 3\_7 and suggests a very patchy reddening
distribution.

\section{Star clusters in STEP}

As discussed in Sect. 2, the study of stellar clusters in the
surveyed area is an important part of the STEP survey. In the previous section we have already 
discussed some of the clusters which can be readily identified in 
tile 4\_6. The clusters (and associations) listed in Table~\ref{crossIdClusters}
represent only part of the cluster/association content of tile
4\_6. Indeed, the compilation by \citet{bic08} reports about 114
objects including both clusters and associations in tile 4\_6
only.  Clearly, most of these objects are small and a significant
number of them could actually  be asterisms. 
A detailed analysis of these objects is beyond the
scope of present paper, and will be object of future works. 
Here, to give a flavour of the survey capabilities in such
kind of studies, we present a more detailed
analysis of two known clusters lying in tile 4\_6, namely NGC\,419 and IC\,1624.

\subsection{Surface brightness profiles of NGC\,419 and IC\,1624}

The number density and surface brightness profiles (SBP) are useful
tools to study the properties of star clusters in different galactic 
environments. These profiles contain information about  the cluster's 
formation and evolution due to internal dynamic process 
and interaction with the galactic environment \citep[see,
e.g.][and references
therein]{Djorgovski1994,Gnedin1997,Binney1998,Lamers2005,Mackey2005}. 

The most commonly used analytical functions to describe the SBPs of star
clusters are those by \citet{King1966} and by
\citet[][]{Elson87,Elson99}, i.e. the so called EFF
profiles (similar to the \citealt{Plummer1915} model), 
being the latter particularly suited for young massive clusters in the
MCs \citep[see e.g.][and references therein]{Elson99,Carvalho2008}. 
For this reason, in the following we will adopt the EEF profiles
to fit the SBPs of NGC\,419 and IC\,1624.

Given the high level of crowding in the cluster cores (with the
consequent large incompleteness, especially in
the case of NGC\,419) we decided to avoid the use of the number
density profile and to concentrate on the SBP.  
We followed the approach devised by e.g. 
\citet{Hill2006,Carvalho2008}, consisting in building the profile
shape to the integrated light, 
which does not suffer incompleteness. 

\begin{figure*}
\centering 
\includegraphics[width=15cm]{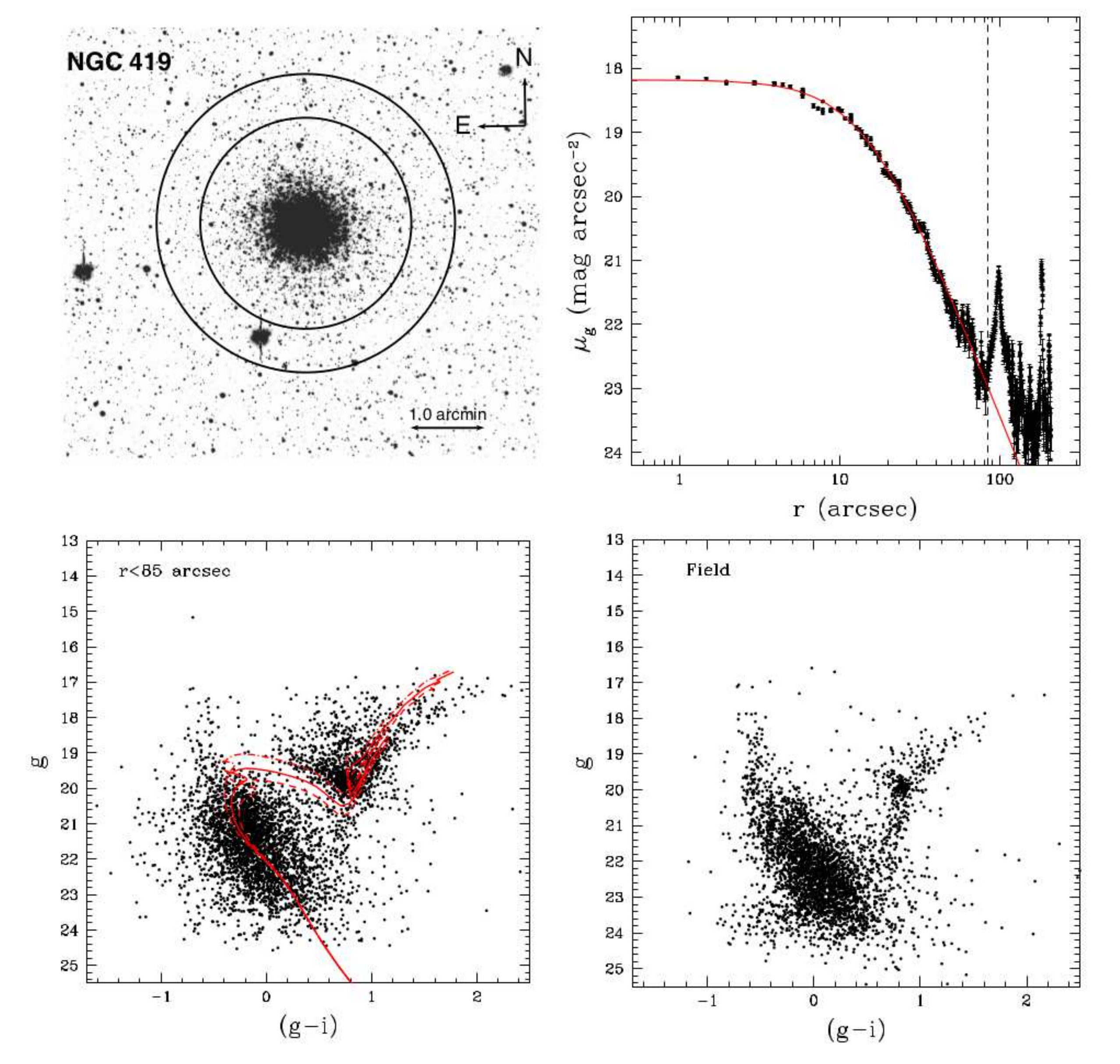}
\caption{Upper-left  panel:  cluster map; the two 
  circles have radii of 85$^{\prime\prime}$ and 120$^{\prime\prime}$,
  respectively.  Upper-right panel: black dots show the observed
  radial stellar surface brightness profile around the NGC\, 419 center.
  The solid red line shows the profile fitting obtained with eq.~\ref{plummer}, while the dashed line represents the
  fitting radius (see text).  Lower-left panel: CMD of the cluster within the
  fitting radius (see text). Lower-right  panel: field stars in an annulus with inner and
  outer radii of 85$^{\prime\prime}$ and 120$^{\prime\prime}$,
  respectively. This annulus has the same area as the  innermost circle. }
\label{figNgc419} 
\end{figure*}

To build the SBP for each cluster we followed closely the recipe 
by \citet{Carvalho2008}, adopting also their coordinates for the cluster's
centres.  Details of the procedure can be found in section 2.2 of \citet{Carvalho2008}. 
Here we recall that  we measured 
the surface brightness in four sets of concentric annuli with radial
steps of 1.5$^{\prime\prime}$,  2.0$^{\prime\prime}$,
3.0$^{\prime\prime}$, and 4.0$^{\prime\prime}$, centred on 
cluster's centres. We took the mean and the standard deviation of the
counts in each annulus. The background was subtracted to the counts,
after evaluating it in a number of loci in the surrounding
of the clusters. The counts were eventually converted to magnitudes
and plotted against the radius from the cluster centre. The result of
such a procedure is shown in the upper-right panel of 
Figures~\ref{figNgc419} and~\ref{figIc1624} for
NGC\,419 and IC\,1624, respectively. 

We then employed the EFF model to analyse the SBPs of the target
clusters. Its analytical formulation in magnitudes is:

\begin{equation}
\mu(r) = \mu(0) +1.25 \gamma \log(1 + r^2/\alpha^2) \\ 
\label{plummer}
\end{equation}

\noindent 
where $\mu(0)$ is the central surface brightness in magnitudes, $\alpha$ is 
a measure of the core radius\footnote{According to the original 
  \citet{Plummer1915} paper, ``The $\alpha$ parameter may be regarded 
  as the radius of the equivalent homogeneous or uniform cluster,
  containing the same number of stars and having the same volume 
  density, or the same areal density, as the actual cluster at its 
  centre.''} and $\gamma$ is the power-law slope. The parameter 
$\alpha$ in the EFF profiles is related to the core radius of a 
cluster by $r_c = \alpha \cdot (2^{2/\gamma} - 1)^{1/2}$. 
Equation~\ref{plummer} is equal to the original \citet{Plummer1915}
function for $\gamma$=4, whereas an empirically derived value suited 
for the (young) MC cluster is 2.7 \citep{Elson87,Carvalho2008}.

 \begin{table}
   \begin{center}
      \caption[]{Results of the stellar density profile fit obtained with  
        Eq.~\ref{plummer}.}
         \label{structure}
         \begin{tabular}{c c  c c c}
            \hline 
            \hline 
          \noalign{\smallskip}
  Cluster & $\mu_g(0)$  & $\alpha$  & $\gamma$  & Fit. radius \\
           \noalign{\smallskip}
              &  mag arcsec$^{-2}$ & arcsec &  & arcsec    \\
           \noalign{\smallskip}
           \hline 
           \noalign{\smallskip}
NGC\,419 & 18.18$\pm$0.01 & 14.8$\pm$0.3 & 2.52$\pm$0.03  & 85  \\ 
IC\,1624 & 19.11$\pm$0.04  & 15.0$\pm$1.5 & 3.2$\pm$0.3  & 39  \\ 
            \noalign{\smallskip}
            \hline
	\end{tabular}
	\end{center}
   \end{table}

We adopted a weighted least-square method to fit Eq.~\ref{plummer} to the
observed SBPs. In order to obtain a better profile, we restricted the
fit to a maximum radius (fitting radius) beyond which the cluster is no
longer distinguishable from the background. We adopt this fitting radius
in the next section as a limit to decide whether or not a star belongs to the cluster. 
The fitting radius is shown as a dashed line in the upper-right panel of 
Figures~\ref{figNgc419} and~\ref{figIc1624}. 
In the same figures, a solid red line shows the result of the fit,
while Tab.~\ref{structure} lists the values of the relevant parameters
of Eq.~\ref{plummer}.

These values compare well with similar analyses present in the
literature. In particular, for cluster NGC\,419, our
results appear to be in good agreement with those by 
\citet{Hill2006,Carvalho2008,glatt09}, taking also into account the
difference in the adopted filter ($g$ instead of $V$), centre
coordinates (apart \citealt{Carvalho2008}), spatial resolution (ccd pixel size), extension (size of the 
frame) and depth of the exposures used in these different investigations.
As for IC\,1624, the agreement with \citet{Carvalho2008} is worse,
(especially for  $\mu$(0)); this is perhaps due to the different
statistics adopted \citep[see Sec. 2.2 and Tab. 3
in][]{Carvalho2008}.

\subsection{The age of NGC\,419 and IC\,1624}

\begin{figure*}
\centering 
\includegraphics[width=15cm]{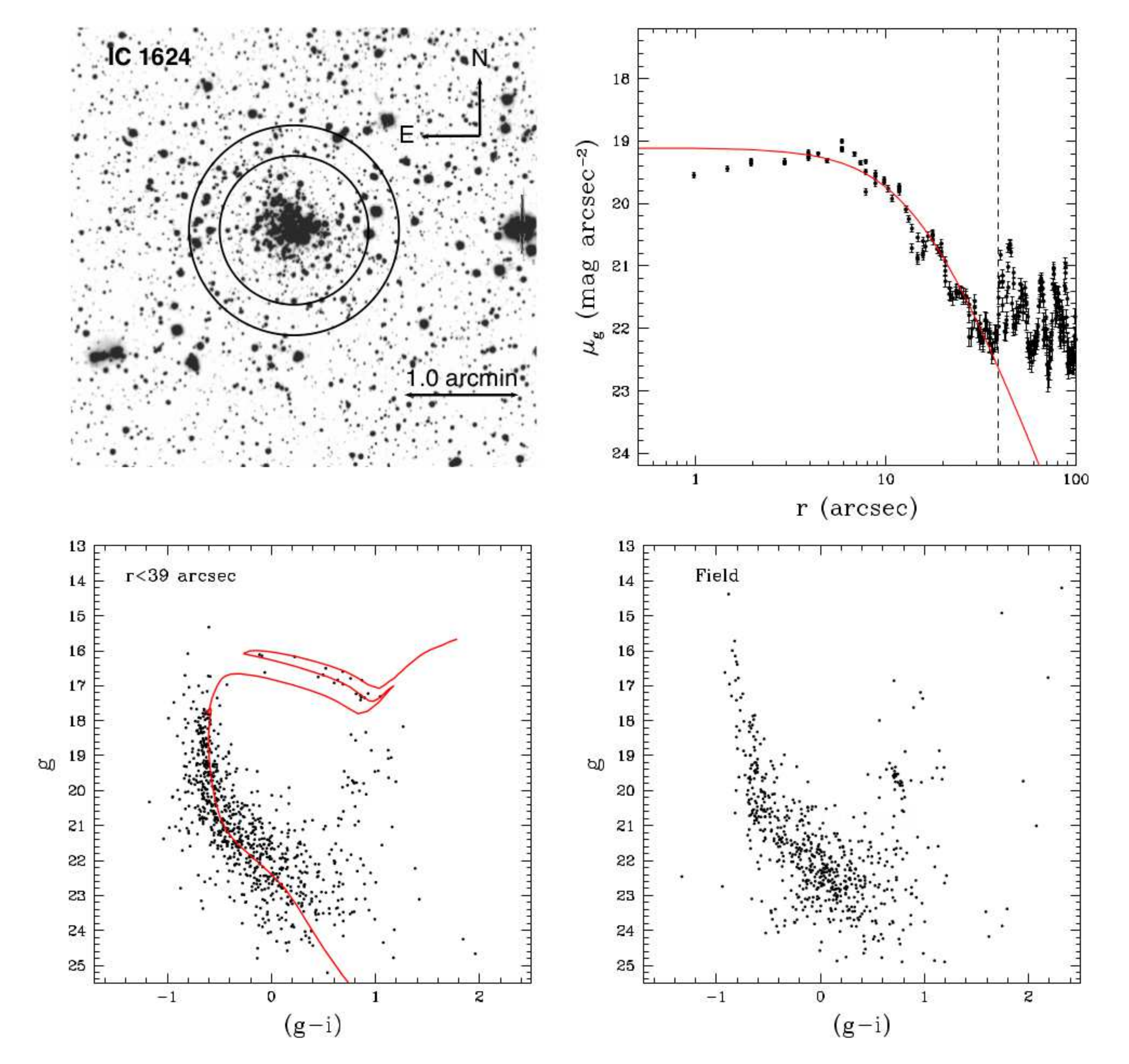}
\caption{Same as in Fig.~\ref{figNgc419}, but for cluster IC\,1624. In
  this case inner and outer circles have radii of 39$^{\prime\prime}$
  and 55$^{\prime\prime}$, respectively.}
\label{figIc1624} 
\end{figure*}

As mentioned in the previous section, we assumed  
the fitting radius to be a (rough) 
measure of the ``effective'' cluster radius (i.e. we consider all  
stars within this radius as belonging to the cluster). 
The soundness of our  choice 
 is confirmed by looking at the upper-left 
panels of Figures~\ref{figNgc419} and~\ref{figIc1624},  even if some 
contamination from the field is still present (see lower-panels of
Fig.~\ref{figIc1624}). Furthermore, the
fitting radii estimated here agree very well with (half of) the major and
minor axes \citep[according to][]{bic08} listed in Tab.~\ref{crossIdClusters}
for the two clusters.

The CMDs of NGC\,419 and IC\,1624, as well as 
those of the surrounding fields are shown
 in the lower-left and lower-right panels of Figures~\ref{figNgc419}
and~\ref{figIc1624}, respectively. The CMD of NGC\,419 is rather scattered, due to the high
crowding, which
  boosts dramatically photometric errors and incompleteness towards
  the cluster centre. Indeed, the CMD within the effective radius is
  almost 1 mag shallower than the CMD of the external field. On the
  other hand, the CMD of IC\,1624 is rather loose because of the
  global paucity of stars in this cluster.

In terms of age, there is a clear difference between the two
clusters, with IC\,1624 being much younger than NGC\,419 (see also
Tab.~\ref{crossIdClusters}). To be more
quantitative, we adopted the isochrones by \citet{marigo08} to
estimate the ages of the two clusters. The best-fitting models are
shown in Figures~\ref{figNgc419} and~\ref{figIc1624} (lower-left panels)
whereas the age and the other relevant parameters are listed in 
Table~\ref{age} (we recall here that the calibration, especially
for blue colours is subject 
to improvement,  which can affect e.g. the reddening estimate in Table~\ref{age}). 

NGC\,419 is known to host multiple stellar populations. A visual
inspection of higher resolution data \citep[HST/HRC, see Fig. 42 in
][]{glatt08a} shows clearly a very broad MSTO, more consistent with a
prolonged SF than a single burst population. Indeed, using isochrone
fitting applied to MSTO stars, \citet{glatt08a} found an age spread
between 1.2 and 1.6 Gyr. Afterwards,  \citet{girardi09} reanalysed the
data modelling the CMD with synthetic populations, and concluded that
only an age spread of about $\Delta \log$age$\approx0.15$ dex
allows to reproduce the MSTO and RC morphologies simultaneously. 
In our data, the magnitude difference between RC and bulk of the
  MSTO suggests an age between 900 and 1100 Myr (see the bottom-left
  panel in Fig.~\ref{figNgc419}), but the large photometric errors
  prevent any definitive conclusion about a genuine age spread. In this
  regard, it is also worth of notice the large number of stars just
  above the RC, which are not seen in the external field.
The presence of these object can be naturally
explained in terms of blending due to the severe crowding conditions. 
Anyway, we cannot exclude that the innermost region of the
cluster is harbouring an even younger sub-population (composed by BL
stars).

The situation is completely different in IC\,1624. The 170 Myr old
isochrone (see Fig. \ref{figIc1624}) fits very well all the main
evolutionary phases, including the MSTO (around $g\sim 18$ mag), the red
envelope of the BL ($g-i \sim 0.8$ mag) and the average luminosity of the
loop. Interestingly, the field around the cluster appears to be 
younger than the cluster itself. This is not surprising given the
proximity of the young Nebula-Association NGC 395 (see
Figures~\ref{ngc346a}, ~\ref{maps_f46} and Table~\ref{crossIdClusters}).

 \begin{table}
   \begin{center}
      \caption[]{Results from the isochrone fitting procedure for
        NGC\,419 and IC\,1624 (see text).}
         \label{age}
         \begin{tabular}{c c  c c c}
            \hline 
            \hline 
          \noalign{\smallskip}
  Cluster & $(m-M)_0$  & $E(B-V)$ & Z & Age  \\
           \noalign{\smallskip}
              & mag & mag & &Myr      \\
           \noalign{\smallskip}
           \hline 
           \noalign{\smallskip}
NGC\,419 &   18.80 & 0.04 & 0.004 & 700,900,1100\\ 
IC\,1624 &   19.0 &  0.02   & 0.004  &  170\\ 
            \noalign{\smallskip}
            \hline\\
	\end{tabular}
	\end{center}
   \end{table}

\section{Summary}

The STEP survey is a deep, homogeneous and uniform $g,r,i,H_{\alpha}$
survey of $\sim$72 deg$^2$ across the SMC and the Bridge, as well as 2
deg$^2$ on the Magellanic Strem.  The STEP
observations, based on INAF VST Guaranteed Time Observations, started
on November 2011.

The STEP data will provide, among other things, a detailed history of
star formation across the SMC and the Bridge, and an inventory of
variable stars along eight fields in the Bridge, outside the region
investigated by the OGLE III survey.

This paper presents the STEP survey strategy, the technique adopted
for the data reduction and the first results aimed at 
assessing their scientific quality. 

The data presented here show the potential of the survey in addressing
its main science goals and validate the adopted observing strategy.
In particular, we analysed qualitatively two tiles, namely tiles  4\_6 and
3\_7, centred on the north of the SMC bar and on the Wing,
respectively.  The CMDs of these fields show a wealth of substructures
and a clear separation between the various populations hosted in the
galaxy. It is also clear that our photometry allows us to investigate
in detail the populations producing the oldest MSTO in the galaxy,
i.e. one of the main aim of this survey.  These diagrams will form the
base of the SFH analysis which will be the subject of future papers.

To illustrate  some additional scientific applications of the STEP survey, 
we investigated two stellar clusters falling into the 4\_6 tile,
namely NGC\,419 and IC\,1624. We used the  stellar radial density profiles 
to estimate their structural parameters, whereas the analysis of their
CMDs allowed us to estimate their age by means of isochrone-fitting.

Finally, the STEP survey will be of great importance for the
astronomical community because it represents the optical complement to
the VMC ESO public survey \citep{Cioni11} which is surveying the
Magellanic System in the near infrared $YJK_\mathrm{s}$ bands at a
comparable level of sensitivity.  Similarly, STEP will complement the
HST observations already obtained in the optical bands, that, albeit
immeasurably better for resolution and depth, are nevertheless
confined to small FoVs throughout the SMC.  Finally, we plan to release
to the community the catalogues as well as the reduced images as soon
as they are fully validated.

\section*{Acknowledgments}

We wish to thank our anonymous Referee for his/her competent and useful
review that helped us to improve the paper.

Partial support to this work was provided by the following projects:
PRIN-MIUR 2010 (2010LY5N2T)
``Chemical and Dynamical evolution of the Milky Way and Local Group
galaxies'' (PI F. Matteucci); PRIN-INAF 2011 ``Galaxy evolution with the
VLT Surveys Telescope (VST)'' (PI A. Grado); PRIN-INAF 2011 ``Tracing the formation
and evolution of the Galactic halo with VST'' (PI M. Marconi).

We acknowledge Amata Mercurio and Crescenzo Tortora for helpful
discussions about the CMD contamination caused by faint extragalactic sources.

We thank the VST Center for providing access to its computational
infrastructures.

This research has made use of the SIMBAD database and VizieR catalogue
access tool, operated at CDS, Strasbourg, France

Funding for SDSS-III has been provided by the Alfred P. Sloan
Foundation, the Participating Institutions, the National Science
Foundation, and the U.S. Department of Energy Office of Science. The
SDSS-III web site is http://www.sdss3.org/.
SDSS-III is managed by the Astrophysical Research Consortium for the
Participating Institutions of the SDSS-III Collaboration including the
University of Arizona, the Brazilian Participation Group, Brookhaven
National Laboratory, University of Cambridge, Carnegie Mellon
University, University of Florida, the French Participation Group, the
German Participation Group, Harvard University, the Instituto de
Astrofisica de Canarias, the Michigan State/Notre Dame/JINA
Participation Group, Johns Hopkins University, Lawrence Berkeley
National Laboratory, Max Planck Institute for Astrophysics, Max Planck
Institute for Extraterrestrial Physics, New Mexico State University,
New York University, Ohio State University, Pennsylvania State
University, University of Portsmouth, Princeton University, the
Spanish Participation Group, University of Tokyo, University of Utah,
Vanderbilt University, University of Virginia, University of
Washington, and Yale University.





\end{document}